# Instantons and Meson Correlation Functions in QCD

**Marcus Hutter**

*Sektion Physik der Universität München*
*Theoretische Physik*
*Theresienstr. 37 80333 München*

**Abstract**

Various QCD correlators are calculated in the instanton liquid model in zero-mode approximation and $1/N_c$ expansion. Previous works are extended by including dynamical quark loops. In contrast to the original "perturbative" $1/N_c$ expansion not all quark loops are suppressed. In the flavor singlet meson correlators a chain of quark bubbles survives the $N_c \to \infty$ limit causing a massive $\eta'$ in the pseudoscalar correlator while keeping massless pions in the triplet correlator. The correlators are plotted and meson masses and couplings are obtained from a spectral fit. They are compared to the values obtained from numerical studies of the instanton liquid and to experimental results.

# Contents

# 1 Introduction

The fundamental theories which describe the observed interactions are all gauge theories. They describe gravitation, electroweak and strong interaction. In the quantized version the forces between particles (fermions) are mediated by gauge bosons. QCD describes the strong interaction between quarks and gluons. It is a SU(3) gauge theory with the Lagrangian[1]

$$\mathcal{L} = \frac{1}{4g^2} G^a_{\mu\nu} G^{\mu\nu}_a + \sum_{i=1}^{N_f} \bar{\psi}_i (i\not{D} + im_i)\psi. \quad (1)$$

Taking into account only the light quarks $u$, $d$ and sometimes $s$ and setting their mass to zero one arrives at a theory with only one parameter, the gauge coupling constant $g$, which is actually no parameter because of dimensional transmutation. Despite the fact that QCD looks such simple (e.g. compared to electroweak interaction) it is very hard to solve this theory due to nonperturbative effects. To clarify the role of instantons in QCD let me first give a list of the most important methods used to tackle QCD starting with very general methods applicable to any quantum field theory (QFT) and ending with more specific approaches:

- *Axiomatic field theory:* Wightman/Oswalder&Schrader have stated a set of Minkowskian/Euclidian axioms for vacuum correlators a general QFT should respect (analyticity, regularity, Lorentz invariance, locality, ...). It is clear that the theorems derived from these axioms have to be very general because no Lagrangian is used [2].

[1] Throughout the whole work the euclidian formulation of QCD is used [11].

- *Haag-Ruelle/LSZ Theory:* S-Matrix elements are related to vacuum correlators. The S-Matrix is the central object for particle phenomenology containing such important informations as cross sections, form factors, structure functions, .... For theoretical studies vacuum correlators are more suitable because they avoid the need of explicitly constructing the Hilbert space which is an extremely complicated task beyond perturbation theory.

- *Quantization:* One might think that quantization should not appear in a list of methods for solving QFT because it is the method to obtain and define a QFT. On the other hand, besides canonical quantization there are other ways of quantizing a theory. The most popular is the path integral quantization [8]. Usually in textbooks for particle physics it is only used as an abbreviation to derive theorems more quickly. In general (beyond perturbation theory), different quantization methods lead to different physical and mathematical insights and different methods to solve the theory. Variants of the path integral quantization are the random walk quantization used in lattice theories and stochastic quantization.

- *(Broken) Symmetry:* Every degree of freedom like spin, flavor and color is the possible origin of (approximate) symmetries like $SU(2N_f)$ or subgroups and $SU(3)$ gauge invariance. Conserved currents and Ward identities [4] can be obtained. In the case of light quarks, one further has approximate chiral symmetry leading to PCAC, axial Ward identities, current algebra theorems, soft pion physics,.... Furthermore, QCD with massless quarks possesses an anomalously broken scale invariance which is the origin of the huge field of renormalization group techniques [5].

- *Perturbation theory:* Due to asymptotic freedom the coupling constant $g$ decreases for high energies and perturbation theory in $g$ is applicable. QCD can thus be solved for processes which involve only momenta of say more than 1 GeV. The small distance behaviour of vacuum correlators is thus calculable ($x \leq 0.2$ fm).

- *Operator Product Expansion (OPE):* An improvement of perturbation theory is to separate the small distance physics from large distance effects. The former is contained in the so-called Wilson coefficients calculated perturbatively. The latter nonperturbative effects are contained in a few vacuum or hadron expectation values of *local* operators which have to be determined from phenomenology or uncertain assumptions [11]. Vacuum correlators can be obtained up to distances of $x \leq 0.3 \ldots 0.5$ fm.

- *QCD Sum rules:* QCD sum rules are widely used to determine hadron masses and couplings. The general method is to assume the existence of certain hadrons and take a resonance+continuum ansatz in the Minkowskian region for some correlator. The Euclidian correlator is calculated from theory (OPE, lattice, instantons). Via dispersion relations one may match both in some Euclidian window by fitting the hadron parameters which leads to a prediction for them.

- *Effective Theories:* One may construct effective Lagrangians containing mesons and/or baryons more or less motivated by QCD or history or other physical branches.

Their parameters are determined from experiment or QCD as far as possible.

A variety of phenomena have been explained qualitatively and calculated quantitatively with the methods listed above, but the large distance behaviour of QCD is still unsolved. There are at least two problems belonging to this domain: chiral symmetry breaking and confinement. Up to now chiral symmetry breaking was assumed and the consequences such as Golstone bosons were discussed within this assumption. In OPE one takes the nonzero values of the quark and other condensates from experiment but has no possibility to predict them from theory. The quark condensate is the order parameter of chiral symmetry and a nonzero values indicates spontaneous breaking of this symmetry (SBCS). Confinement has also to be assumed.

These two problems are not so unsolved as it might appear. Up to now I have not mentioned two further approaches:

- *Lattice QCD:* In principle the method is very simple. The continuum is replaced by a fine lattice covering a large but finite volume. The path integral is thus replaced by a finite number of integrals evaluated numerically. All vacuum correlators can be obtained for arbitrary Euclidian distances. Confinement and SBCS have been shown and other hadronic parameters are obtained. In practice, lattice calculations are much less straightforward than this scetchy description might suggest [9].

- *Instantons:* As in lattice QCD one evaluates the Euclidian path integrals but now in semiclassical approximation. In addition to the global minimum of the QCD action $A^a_\mu = 0$ used by perturbation theory there are many other local minima called instantons which have to be taken into account. Inclusion of light quarks leads to an effective $2N_f$ quark vertex responsible for SBCS. Although confinement cannot be explained, a lot of hadron parameters can nevertheless be calculated [20, 21] suggesting that confinement is not essential for the properties of hadrons.

In this article I want to give a quantitative and systematic study of the implications of instantons to hadron properties.

Chapter 2 is an introduction to the semiclassical evaluation of nontrivial integrals. After developing the method in the finite dimensional case the partition function of QCD is considered and the instanton liquid model is introduced.

In chapter 3 the propagator of a light quark is calculated. The approximations which have to be made are stated and discussed. It is shown that for one quark flavor the $1/N_c$ expansion is exact. The constituent quark masses and the quark condensates are calculated for $u$,$d$ and $s$ quarks.

The same approximations are used in chapter 4 to calculate the 4 point functions. Special attention is payed to the singlet correlator where a chain of quark loops contributes and is *not* suppressed in the large-$N_c$ limit. Within one and the same approximation we get Goldstone bosons in the pseudoscalar triplet correlator but no massless singlet boson.

In chapter 5 meson correlators are discussed and plotted. Employing a spectral ansatz it is possible to extract various meson masses and couplings. They are compared to the values obtained from extensive numerical studies of the instanton liquid [21] and to the experimental values.

Conclusions are given in chapter 6.

# 2 Theory of Instanton Liquid

### Separating Gaussian from non-Gaussian degrees of freedom

In the path integral formulation solving QCD or any other QFT is equivalent to calculating the partition function $Z$ including external source terms. Of course this is not a simple task because one has an infinite number of degrees of freedom. One way to tackle this problem is to reduce the number of degrees of freedom by integrating the (nearly) gaussian degrees perturbatively leaving the non-Gaussian degrees, called collective coordinates, to be treated with other methods.

Let me first describe this method most generally without refering to QFT. Our aim is to calculate

$$Z = \int d^n x \, e^{-S[x]} \tag{2}$$

where S is some real valued function depending on the $n$ dimensional vector $x \in I\!R^n$. Let the integral be dominated by values of $x$ which lie in the vicinity of a $k$ dimensional sub-manifold of $I\!R^n$ which may be parametrisized by $x = f(\gamma)$, $\gamma \in I\!R^k$. Vectors in the tangent space of $f(\gamma)$ are called (approximate) zeromodes, because $S[x(\gamma)]$ is (nearly) constant in this direction. Usually $f(\gamma)$ represents some degenerate or approximate minima of $S$. In addition $f(\gamma)$ may contain points which do not contribute much to the functional integral, this will cause no error, but $f(\gamma)$ must not forget any significant points. In figure 1 a two dimensional example is shown containing a river (one dimensional manifold) with steep mountains aside his banks. So the integral will be dominated by the shaded area. This area can be parametrized in the following way:

$$x = f(\gamma) + y \quad , \quad x, y \in I\!R^n \quad , \quad \gamma \in I\!R^k \quad . \tag{3}$$

To make this representation unique, we have to demand $k$ linear extra conditions to $y$:

$$y \cdot g_i(\gamma) = c_i \quad , \quad 1 \leq i \leq k \tag{4}$$

E.g., $c_i = 0$ and $g_i(\gamma) = \partial f(\gamma)/\partial \gamma_i$ would fix $y$ to be orthogonal to the "river" or equally stated: It disallows fluctuations in the non-Gaussian zeromode direction. Now every point $x$ (at least in the vicinity of the river) can uniquely be described by $y$ satisfying the condition (4) and by $\gamma$ via (3). All we have to do now is to represent $Z$ in terms of the fluctuation vector $y$ and the collective coordinates $\gamma$. A convenient way is to introduce a Faddeev-Popov unit

$$1 = \int d^k\gamma \, d^n y \, \delta^k(y \cdot g_i(\gamma) - c_i) \, \delta^n(f(\gamma) + y - x) \, \Phi(x) \tag{5}$$

which serves as a definition of $\Phi$. Inserting this into (2) and integrating over $x$ yields

$$Z = \int d^k\gamma\, d^n y\, \delta^k(y\cdot g_i(\gamma) - c_i)\, \Phi(f(\gamma)+y)\, e^{-S[f(\gamma)+y]} \quad . \tag{6}$$

The $y$-integration in (5) can trivially be performed:

$$\Phi^{-1}(x) = \int d^k\gamma'\, \delta^k((x - f(\gamma'))g_i(\gamma') - c_i) \quad . \tag{7}$$

For $x = f(\gamma) + y$ the $\delta$-function in the integral only contributes for $\gamma' = \gamma$. Thus we can expand the $\delta$-argument up to linear order in $\gamma' - \gamma$

$$\begin{aligned} \Phi^{-1}(f(\gamma)+y) &= \int d^k\gamma' \delta^k \left( \sum_j (y\cdot \tfrac{\partial g_i(\gamma)}{\partial \gamma_j} - g_i(\gamma)\cdot \tfrac{\partial f(\gamma)}{\partial \gamma_j})(\gamma'_j - \gamma_j) \right) \\ &= \left|\det_{ij}\,(y\cdot\partial_j g_i - g_i\cdot\partial_j f)\right|^{-1} \quad {}^{2} \end{aligned} \tag{8}$$

$$\text{with} \quad y\cdot g_i(\gamma) = c_i \quad .$$

Inserting $\Phi$ into (6) we get

$$Z = \int d^k\gamma\, d^n y\, \delta^k(y\cdot g_i(\gamma) - c_i) \det_{ij}\,(y\cdot\partial_j g_i(\gamma) - g_i(\gamma)\cdot\partial_j f(\gamma))\; e^{-S[f(\gamma)+y]} \quad . \tag{9}$$

One may write $Z$ in a slightly different form, usually used in QFT, because it is more suitable for semiclassical approximations. $Z$ is independent of $c_i$ and therefore, although the r.h.s. of (9) explicitly contains $c_i$, it is actually independent of it. We can smooth the $\delta$-function by a further multiplication with

$$1 = (2\pi\xi)^{-k/2} \int d^k c\, e^{-\frac{1}{2\xi}\sum_i c_i^2} \quad . \tag{10}$$

The determinant can be written in the form

$$\det A = \int d\eta\, d\bar\eta\, e^{\bar\eta A \eta} \tag{11}$$

where $\eta$ are anticommuting grassmann variables (ghosts). Inserting (10) and (11) into (9) and performing the $c$-integration we finally get

$$\begin{aligned} Z &= (2\pi\xi)^{-k/2} \int d^k\gamma\, d^n y\, d\eta\, d\bar\eta\, e^{-S[f(\gamma)+y] - S_{gf}[y,\gamma] + S_{FPG}[y,\gamma,\eta,\bar\eta]} \quad , \\ S_{gf} &= \frac{1}{2\xi} y^T \left( \sum_i g_i(\gamma) g_i^T(\gamma) \right) y \quad , \\ S_{FPG} &= \sum_{ij} \bar\eta_i\,(y\cdot\partial_j g_i(\gamma) - g_i(\gamma)\cdot\partial_j f(\gamma))\,\eta_j \quad . \end{aligned} \tag{12}$$

---

[2] In the following we will omit the absolute bars; when necessary they have to be reinstated.

**Effective QCD Lagrangian in a background field**

Now it is time to return to QCD

$$Z = \int DA_\mu \, e^{-S_{YM}[A]} \quad , \quad S_{YM}[A] = \int dx \, \frac{1}{4g^2} G^a_{\mu\nu} G^{\mu\nu}_a \quad . \tag{13}$$

The background configurations which (approximately) minimize $S_{YM}$ will be denoted by $\bar{A}_\mu(\gamma)$. A general gauge field can be written in the form

$$A_\mu = (\bar{A}_\mu(\gamma) + B_\mu)^\Omega \quad , \quad \gamma \in I\!R^k \tag{14}$$

where $B_\mu$ are fluctuations around the background and $A^\Omega_\mu$ is a gauge transformed field

$$A^\Omega_\mu = S A_\mu S^\dagger + i S \partial_\mu S^\dagger \quad , \quad S = e^{i\Omega} \in SU(N_c) \quad . \tag{15}$$

As in the finite dimensional case we have to make this representation unique by introducing extra conditions,

$$D_\mu(\bar{A}) B_\mu(x) = C(x) \quad , \tag{16}$$

to fix the gauge of the fluctuating field and

$$\int d^4x \, \psi^i_\mu(x;\gamma) B_\mu(x) = c_i \quad , \quad 1 \le i \le k \tag{17}$$

to avoid fluctuations in (approximately) zero mode direction. The derivation of an effective action similar to (12) can now be performed in full analogy to the previous case with only some notational complication. There is the following correspondence:

$$\begin{array}{lcccccc} \text{finite example} & : & i & x & y & \gamma_i & g_i \\ \text{QCD} & : & i, x & A & B & \gamma_i, \Omega(x) & \psi^i_\mu \end{array} \quad . \tag{18}$$

The Faddeev-Popov unit has the form

$$1 = \int d^k\gamma \, D\Omega \, DB_\mu \, \delta(D_\mu(\bar{A})B_\mu) \delta^k(\int \psi^i_\mu B_\mu d^4x) \delta((\bar{A}_\mu + B_\mu)^\Omega - A_\mu) \Phi[A_\mu] \quad . \tag{19}$$

The steps to get an expression for $\Phi$ are now:

Add primes to $\gamma, \Omega$ and $B$, insert $A = \bar{A} + B$, linearize the last $\delta$-argument around $B'_\mu = B_\mu$, perform the functional $B'_\mu$ integration and linearize the remaining $\delta$ arguments around $\gamma' = \gamma$ and $\Omega' = 0$. Omitting the details of this calculation one gets [19]

$$\Phi^{-1}(\bar{A}(\gamma) + B) = \int d^k\gamma' \, D\Omega' \delta^k(X_i) \, \delta(Y) \quad , \tag{20}$$

$$X_i = \int d^4x \sum_j \left( \psi^i_\mu(\gamma) \frac{\partial \bar{A}_\mu}{\partial \gamma_j} - \frac{\partial \psi^i_\mu(\gamma)}{\partial \gamma_j} B_\mu \right) (\gamma'_j - \gamma_j) + \psi^i_\mu D_\mu(\bar{A} + B)\Omega' \quad , \tag{21}$$

$$Y = \sum_j D_\mu(\bar{A}+B)\frac{\partial \bar{A}_\mu}{\partial \gamma_j}(\gamma_j' - \gamma_j) + D_\mu(\bar{A})D_\mu(\bar{A}+B)\Omega' \quad . \tag{22}$$

From (16), (17), (21) and (22) one can read off the form of the partition function $Z$:

$$Z = N(\xi)\int d^k\gamma\, DB_\mu\, D\eta\, D\bar{\eta}\delta^k(\int \psi^i_\mu B_\mu d^4x)e^{-S_{QCD}[\bar{A},B,\eta,\bar{\eta}]} \quad , \tag{23}$$

$$\begin{aligned}
S_{QCD} &= S_{YM}[\bar{A}+B] - S_{gf}[\bar{A},B] + S_{FPG}[\bar{A},B,\eta,\bar{\eta}] \quad , \\
S_{YM} &= \int d^4x \frac{1}{4g^2} G^a_{\mu\nu} G^{\mu\nu}_a(\bar{A}+B) \quad , \\
S_{gf} &= \frac{1}{2\xi}\int d^4x\, (D_\mu(\bar{A})B_\mu)^2 \quad , \\
S_{FPG} &= \sum_{ij}\bar{\eta}_i\left[\int d^4x\, \psi^i_\mu(\gamma)\frac{\partial \bar{A}_\mu}{\partial\gamma_j} - \frac{\partial \psi^i_\mu(\gamma)}{\partial \gamma_j}B_\mu\right]\eta_j \\
&+ \sum_i \int d^4x\, \bar{\eta}_i\psi^i_\mu D_\mu(\bar{A}+B)\eta(x) + \int d^4x \sum_j \bar{\eta}(x)D_\mu(\bar{A}+B)\frac{\partial \bar{A}_\mu}{\partial \gamma_j} \\
&+ \int d^4x\, \bar{\eta}(x)D_\mu(\bar{A})D_\mu(\bar{A}+B)\eta(x) \quad .
\end{aligned} \tag{24}$$

$S_{QCD}$ does not depend on the gauge parameter $\Omega$. For this reason the $\Omega$ integration can be absorbed in the normalization factor $N(\xi)$. $\eta(x)$ are the usual ghost fields originating from the gauge fixing. For every extra condition (17) one gets an additional ghost variable $\eta_i$. For $\bar{A}=0$ and no extra condition ($k=0$) the action given above just reduces to the usual QCD action including Faddeev-Popov ghosts in $R_\xi$ gauge

$$S_{gf} = \frac{1}{2\xi}\int d^4x\, (\partial_\mu B_\mu)^2 \quad , \quad S_{FPG} = \int d^4x\, \bar{\eta}(x)\partial_\mu D_\mu(B)\eta(x) \quad . \tag{25}$$

Note that the action (23) is still exact with the non-harmonic degrees of freedom $\gamma_i$ now separated from the hopefully more gaussian ones, $B_\mu$ and $\eta$.

For small coupling $g$ it is now possible to establish feynman rules from (24) in analogy to the case with no background. For this one has to know the "free" gluon, ghost and quark propagator in a given background $\bar{A}$. If $\bar{A}$ is a non constant field even this is a very complicated task in contrast to usual perturbation theory around $\bar{A}=0$. For a multi-instanton configuration explicit expressions for the gluon and ghost propagator are derived in [16, 24] and for light quark propagators in chapter 3.

### The Semiclassical Limit

Before developing perturbation theory to all orders it is wise to study the semiclassical limit where one keeps only terms up to quadratic order in the fields. In QCD (and many other field theories) this is equivalent to lowest order perturbation theory, but around a very nontrivial background!

Up to now we have not specified $\psi^i_\mu$. A natural choice would be $\psi^i_\mu = \partial \bar{A}_\mu / \partial \gamma_i$ to fix the fluctuations to be orthogonal to the zero modes. Somewhat more convenient is to bring $\psi^i_\mu$ in background gauge:

$$\psi^i_\mu = \left( \frac{\partial \bar{A}_\mu}{\partial \gamma_j} \right)^\Omega \quad \text{with } \Omega \text{ such that} \quad D_\mu(\bar{A})\psi^i_\mu = 0 \quad . \tag{26}$$

Furthermore we assume that $\bar{A}$ minimizes the gauge action $S_{YM}$ which is true for widely separated instantons thus neglecting linear terms $S_{QCD}$.

Up to quadratic order in the fields one has

$$\begin{aligned} S_{QCD} &= S_{YM}[\bar{A}] + \int d^4x \, \frac{1}{2g^2} B_\mu K_{\mu\nu}(\bar{A}) B_\nu + \int d^4x \, \bar{\eta}(x) D^2(\bar{A}) \eta(x) \\ &+ \sum_{ij} \bar{\eta}_i \psi^i_\mu \frac{\partial \bar{A}_\mu}{\partial \gamma_j} \eta_j + \int d^4x \sum_j \bar{\eta}(x) D_\mu(\bar{A}) \frac{\partial \bar{A}_\mu}{\partial \gamma_j} \eta_j + O(\text{field}^3) \quad , \end{aligned} \tag{27}$$

$$K_{\mu\nu} = -D^2 \delta_{\mu\nu} + 2iG_{\mu\nu} + (1 - \frac{1}{\xi}) D_\mu D_\nu \quad , \quad G_{\mu\nu} = F^c G^c_{\mu\nu} \quad , \quad (F^c)_{ab} = i f_{acb} \quad . \tag{28}$$

Performing the integration over gauge fields and ghosts one gets an effective action depending only on the collective coordinates $\gamma_i$:

$$Z = \int d^k\gamma \, e^{-S_{eff}[\gamma]} \tag{29}$$

$$e^{-S_{eff}[\gamma]} = \det_{ij} \left( \psi^i_\mu(\gamma) \frac{\partial \bar{A}_\mu}{\partial \gamma_j} \right) \frac{\text{Det}(-D^2(\bar{A}))}{(\text{Det}' K'_{\mu\nu}(\bar{A}))^{1/2}} e^{-S_{YM}[\bar{A}]} \tag{30}$$

The $\delta$-function in (23) causes a restriction of the gauge field fluctuation to be orthogonal to $\psi^i_\mu$. $K'_{\mu\nu}$ is defined as $K_{\mu\nu}$ projected to the space orthogonal to $\psi^i_\mu$, $\text{Det}'$ takes into account all eigenvalues of $K'_{\mu\nu}$ except the $k$ zeromodes caused by the projection.

### The Instanton Gas Approximation

For $N$ well separated pseudoparticles $\bar{A}(x) = \sum A_I(x)$ the partition function $Z$ factorizes like

$$Z = \frac{1}{N!} \prod_{I=1}^{N} Z_I \tag{31}$$

where $Z_I$ is the partition function in a one pseudoparticle background, i.e. with $A_I$ inserted in (30) instead of $\bar{A}$. It was a great deal to evaluate the functional determinants in the background of one instanton[3]

$$\begin{aligned} A^a_{I\mu}(x) &= O_I^{ab}\eta^I_{b\mu\nu}\frac{(x-z_i)_\nu}{(x-z_i)^2}\frac{2\rho^2}{(x-z_I)^2+\rho^2} \quad , \qquad (32)\\ \gamma_I=(z_I,O_I,\rho_I,Q_I) &= (\text{location, orientation, radius, topological charge}) \quad . \end{aligned}$$

To get finite results $Z_I$ has to be normalized to the $\bar{A}^a_\mu = 0$ case, regularized and renormalized. The final result of the very complicated calculation is [14, 15]

$$\begin{aligned} \left(\frac{Z_I}{Z_0}\right)_{reg} &= \frac{1}{2}\sum_{Q_I=\pm}\int d^4z_I dO_I d\rho_I\, D(\rho_I) = V_4\int_0^\infty d\rho\, D(\rho) \quad , \\ D(\rho) &= \frac{1}{\rho^5}\frac{4.6e^{-1.679N_c}}{\pi^2(N_c-1)!(N_c-2)!}S_0(\rho)^{2N_c}e^{-S_0(\rho)} \quad , \qquad (33)\\ S_0(\rho) &= \frac{8\pi^2}{g^2(\rho)} = b\ln\frac{1}{\rho\Lambda}+\frac{b'}{b}\ln\ln\frac{1}{\rho\Lambda}+O\Big(\frac{1}{\ln\frac{1}{\rho\Lambda}}\Big) \quad , \quad \Lambda=\Lambda_{PV} \quad . \end{aligned}$$

$D(\rho)$ is the instanton density, $g(\rho)$ the running coupling constant, $b=\frac{11}{3}N_c$ and $b'=\frac{17}{3}N_c^2$.

**Quarks**

Additional fields coupled in a gauge invariant way to the gluon field can simply be incorporated by adding the appropriate lagrangian with gauge field $A$ replaced by $\bar{A}+B$ and performing the functional integration over the new fields. So every quark contributes an extra factor

$$\int D\Psi D\bar{\Psi} e^{-\int dx\,\bar{\Psi}(i\not{D}+im)\Psi} = \mathrm{Det}(i\not{D}+im) \qquad (34)$$

to the partition function $Z$ where

$$iD_\mu = i\partial_\mu + \bar{A}_\mu + B_\mu \qquad (35)$$

is the covariant derivative. In the semiclassical approximation $B_\mu$ can be set to zero. $D(\rho)$ has to be multiplied by the fermionic factor

$$F(m\rho) = \begin{cases} 1.34m\rho(1+m^2\rho^2\ln(m\rho)+\ldots) & \text{for} \quad m\rho \ll 1 \\ 1-\frac{2}{75m^2\rho^2}+\ldots & \text{for} \quad m\rho \gg 1 \end{cases} \qquad (36)$$

and $b$ and $b'$ are now

$$b=\frac{11}{3}N_c-\frac{2}{3}N_f \quad , \quad b'=\frac{17}{3}N_c^2-\frac{13}{3}N_cN_f+\frac{1}{2}\frac{N_f}{N_c} \quad . \qquad (37)$$

[3] To simplifiy notations we will treat instantons and anti-instantons on the same footing. Both will be called instantons and are distinguished by their topological charge $Q_I=\pm$ if necessary.

**The Instanton Liquid Model**

The probabiliy of small size instantons is low because $D(\rho)$ vanishes rapidly for small distances. On the other hand for large distances $D(\rho)$ blows up and soon gets large. This is the origin of the infrared problem which made a lot of people no longer believing in instanton physics. Those who were not deterred by that have thought of the following outcome [20]. For larger and larger distances, the vacuum gets more and more filled with instantons of increasing size. At some scale the instanton gas approximation breaks down and one has to consider the interaction between instantons which might be repulsive to stabilize the medium. The stabilization might occur at distances at which a semiclassical treatment is still possible and at densities at which the various instantons are still well separated objects - say - not much deformed through their interaction. So there is a narrow region of allowed values for the instanton radius. This picture of the vacuum is called the instanton liqiud model. The idea has been confirmed in the course of years by very different approaches:

- Hardcore assumption [18]
- Variational Approach [19]
- Numerical studies [20]
- Phenomenological success [21]

The picture has now become generally accepted at least by those who believe in instanton physics and it seems that the vacuum can be described by effectively independent instantons of size $\rho = 600\,\mathrm{MeV}^{-1}$ and mean distance $L_0 = 200\,\mathrm{MeV}$. The integral instanton density is fixed by the experimentally known gluon condensate [25]:

$$n = N/V_4 = 1/L_0^4 = \frac{1}{32\pi^2} < G^a_{\mu\nu} G^{\mu\nu}_a >= (200\,\mathrm{MeV})^4_{exp.} \tag{38}$$

The ratio $L_0/\rho$ is estimated in different works to be

$$(L_0/\rho)_{theor.} = 3.0 \ldots 3.2 \tag{39}$$

# 3 Light Quark Propagator

In this chapter the average quark propagator in the multiinstanton background will be calculated. In the first section the instanton background is treated as a classical external perturbation, but the background field is not small (e.g. in the coupling $g$) and therefore we have to sum up *all* feynman graphs. This is possible in the case of one quark flavour within the so-called zeromode approximation. The quark condensate and a constituent quark mass are extracted from the quark propagator. In the last section it is shown that the case of two or more quark flavours can be reduced to the one flavour case in the limit of $N_c \to \infty$. The results obtained in the one flavour case are therefore still valid when making this further approximation.

**Perturbation Theory in the Multi-instanton Background**

It is well known how to calculate correlators in the presence of an external classical gauge field at least as perturbation series in powers of the external field $A_\mu^a(x)$. In the case of QCD (or more accurately in classical chromodynamics) within the instanton liquid model the external field is a sum of well separated scatterers $A = \sum_I A_I$ called instantons with a fixed radius $\rho$ and distributed randomly and independently in Euclidian space.

For a while we will restrict ourself to the case of one quark flavor and ignore gluon loops. The Euclidian feynman rules have the following form:

$$= \frac{1}{\not p + im} = S_0 \quad ,$$

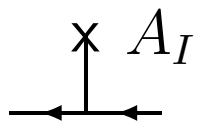

$$= \not A_I \quad . \tag{40}$$

In operator notation,

$$\langle p|S_0|q\rangle = \frac{1}{\not p + im}\bar\delta(p-q) \quad , \quad \langle x|\not A_I|y\rangle = \not A_I(x)\delta(x-y) \quad , \tag{41}$$

$$\bar\delta^d(\cdots) := (2\pi)^d\delta(\cdots) \quad , \quad \int \bar d^d p := \int \frac{d^d p}{(2\pi)^d} \quad ,$$

graphs are simply alternating chains of $S_0$ and $A_I$. To average a graph over the instanton parameters one has to perform the following integration for each instanton:

$$\langle \ldots \rangle_I = \frac{N}{V_4}\int d\gamma_I \ldots = \frac{N}{2}\sum_{Q_I=\pm}\frac{1}{V_4}\int d^4 z_I \int dO_I \ldots \quad . \tag{42}$$

For example

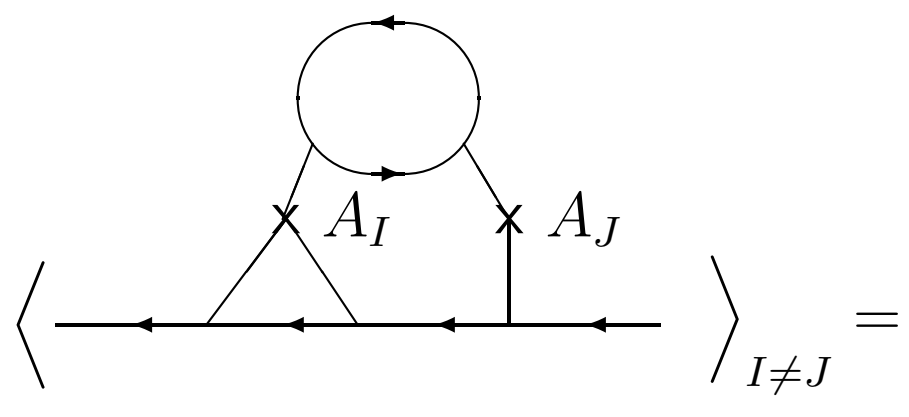

$$= -\frac{N^2}{V_4^2}\int d\gamma_I d\gamma_J \langle p|S_0\not A_I S_0 \not A_I S_0 \not A_J S_0|q\rangle \mathrm{Tr}(S_0\not A_J S_0 \not A_I) \quad . \tag{43}$$

The origin of the factor $N^2$ is the summation over all pairs of different instantons $(I, J)$ yielding a factor $N(N-1) \approx N^2$. The quark loop is the origin of the minus sign and of the functional trace "Tr".

**Exact Scattering Amplitude in the one Instanton Background**

Perturbation theory is suitable to study scattering processes. To achieve chiral symmetry breaking or bound states one has to sum up infinite series of a subclass of graphs or solve Schwinger Dyson or Bethe Selpeter equations.

The first thing we can do is to sum up successive scatterings at one instanton

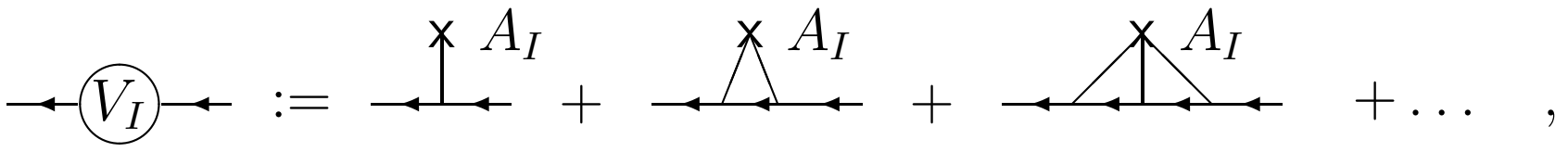

$$V_I := \not{A}_I + \not{A}_I S_0 \not{A}_I + \not{A}_I S_0 \not{A}_I S_0 \not{A}_I + \ldots = S_0^{-1}(S_I - S_0)S_0^{-1} \tag{44}$$

where

$$S_I = (S_0^{-1} - \not{A}_I)^{-1} \tag{45}$$

is the quark propagator in the one instanton background.

**Zero Mode Approximation**

The quark propagator in the one instanton background can be represented in terms of the eigenvalues $\lambda_i$ and eigenfunctions $\psi_i$ of the Dirac operator

$$(i\not{\partial} - \not{A}_I)\psi_i = \lambda_i \psi_i \quad \Longrightarrow \quad \langle x|S_I|y\rangle = \sum_i \frac{\psi_i(x)\psi_i^\dagger(y)}{\lambda_i + im} \quad . \tag{46}$$

There is one zero eigenvalue $i\not{D}\psi_I = 0$ (↑appendix A) which makes the propagator singular in the chiral limit:

$$S_I = \frac{|\psi_I\rangle\langle\psi_I|}{im} + S_I^{NZM} \quad . \tag{47}$$

In the so called zero mode approximation one replaces the non zero mode part by the free propagator

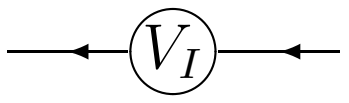

$$= S_I - S_0 \approx \frac{|\psi_I\rangle\langle\psi_I|}{im} \quad . \tag{48}$$

Although this approximation is good for large as well as for small momenta it may be bad for intermediate ones, but what is more important is the fact that it is a wild approximation and so might violate general theorems like Ward identities. In contrast, all other approximations we make are of systematic nature respecting all known symmetries of QCD.

- semiclassical approximation (systematic)
- multi-instanton background ("systematic")
- large $N_c$ expansion (systematic) (see next section)
- zero mode approximation (wild)

Note that every choice of a background gauge field is "systematic" in the sense of respecting the symmetries of QCD as long as the background satisfies these symmetries on average.

In the following sections we will see that the advantage of the zeromode approximation is so great that we cannot disregard this simplification.

**Effective Vertex in the Multi-instanton Background**

Let us consider a quark line with two scatterings at $V_I$ and insert in between a number of instantons which differ from $I$ and from all other instantons occuring elsewhere in the graph. This enables us to average over these enclosed instantons independently from the rest of the graph. Summation over all possible insertions with at least one instanton just yields the exact quark propagator minus the free propagator. Remember that direct repeated scattering at $A_I$ is already included in $V_I$.

Let us define

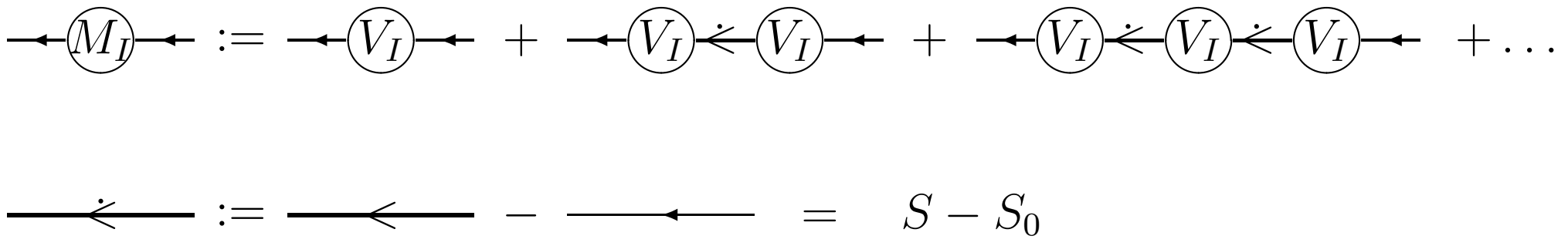

(49)

$$
\begin{aligned}
M_I &= V_I + V_I(S - S_0)V_I + V_I(S - S_0)V_I(S - S_0)V_I + \dots \\
&= V_I + V_I(S - S_0)M_I
\end{aligned}
$$

This equation can be solved for $M_I$ with the following ansatz:

$$M_I = \frac{1}{i\mu} S_0^{-1} |\psi_I\rangle\langle\psi_I| S_0^{-1} \tag{50}$$

Inserting $M_I$ and $V_I$ into (49) we get

$$\frac{1}{i\mu} S_0^{-1}|\psi_I\rangle\langle\psi_I|S_0^{-1} = \frac{1}{im}(1 + \frac{1}{i\mu}\langle\psi_I|S_0^{-1}(S - S_0)S_0^{-1}|\psi_I\rangle)S_0^{-1}|\psi_I\rangle\langle\psi_I|S_0^{-1} \tag{51}$$

$$\Longrightarrow \quad \mu = m + i\langle\psi_I|S_0^{-1}(S - S_0)S_0^{-1}|\psi_I\rangle \quad . \tag{52}$$

**A Nice Cancellation**

It is possible to arrange the graphs in such a way that every $M_I$ occurs only once. Consider a graph containing two scattering processes $M_I$ at the same instanton. The interesting part of the graph has the following form

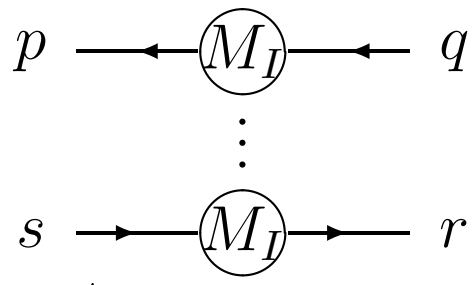

$$= {}^{\alpha_p}_{i_p}\left[\frac{\psi_I(p)\psi_I^\dagger(q)}{i\mu}\right]^{\alpha_q}_{i_q} {}^{\alpha_r}_{i_r}\left[\frac{\psi_I(r)\psi_I^\dagger(s)}{i\mu}\right]^{\alpha_s}_{i_s} \tag{53}$$

$\alpha_p/i_p$ are the color/Dirac indices at the quark leg with momentum $p$. In a physical process in addition to the graph containing the above subgraph there exists another graph with only two quark lines interchanged.

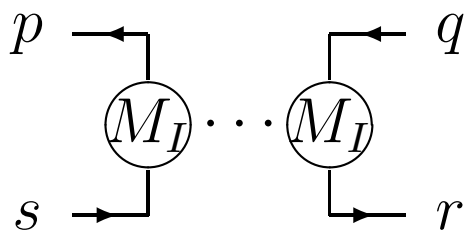

$$= - {}^{\alpha_p}_{i_p}\left[\frac{\psi_I(p)\psi_I^\dagger(s)}{i\mu}\right]^{\alpha_s}_{i_s} {}^{\alpha_r}_{i_r}\left[\frac{\psi_I(r)\psi_I^\dagger(q)}{i\mu}\right]^{\alpha_q}_{i_q} \tag{54}$$

As usual, the interchange of two quark lines causes a minus sign in the amplitude. Inspecting the two expressions, we see that they coincide except for the sign, thus there exists a complete cancellation

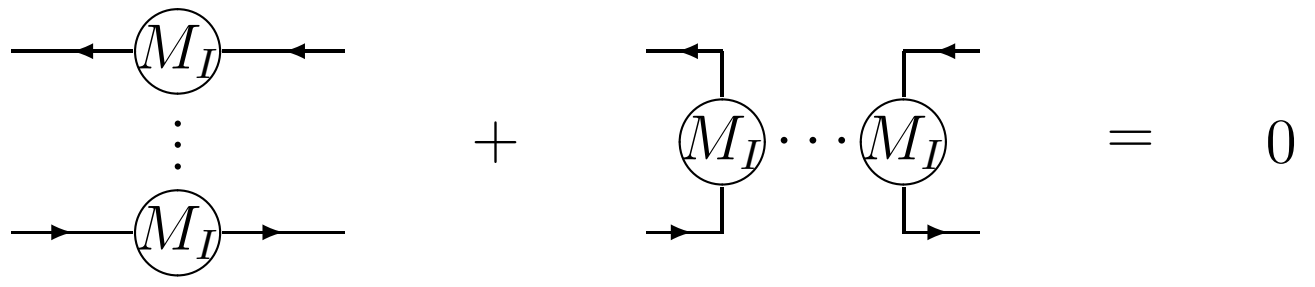

(55)

Whenever an $M_I$ occurs twice or more than twice in a graph there exists another graph with opposite sign. Both contributions cancel each other and can be ignored. So a quark can scatter only once at every instanton. This can be seen in another way: Because of Fermi statistics every state can be occupied only once, and there is only one state for each quark in the zero mode approximation, namely the zeromode.

There are two equivalent descriptions of feynman graphs:

1. Draw all topologically distinct graphs with non-numerated vertices and assign a symmetry factor to each graph,
2. Draw all topologically distinct graphs with numerated vertices and assign a factor $1/V$, where $V$ is the number of vertices.

If all possible graphs containing $M_I$ are allowed it is not difficult to see within the second description that they can really be paired as stated above.

### Renormalization of the Instanton-Density

Up to now the cancellation is incomplete because not all graphs are allowed. Consider e.g.

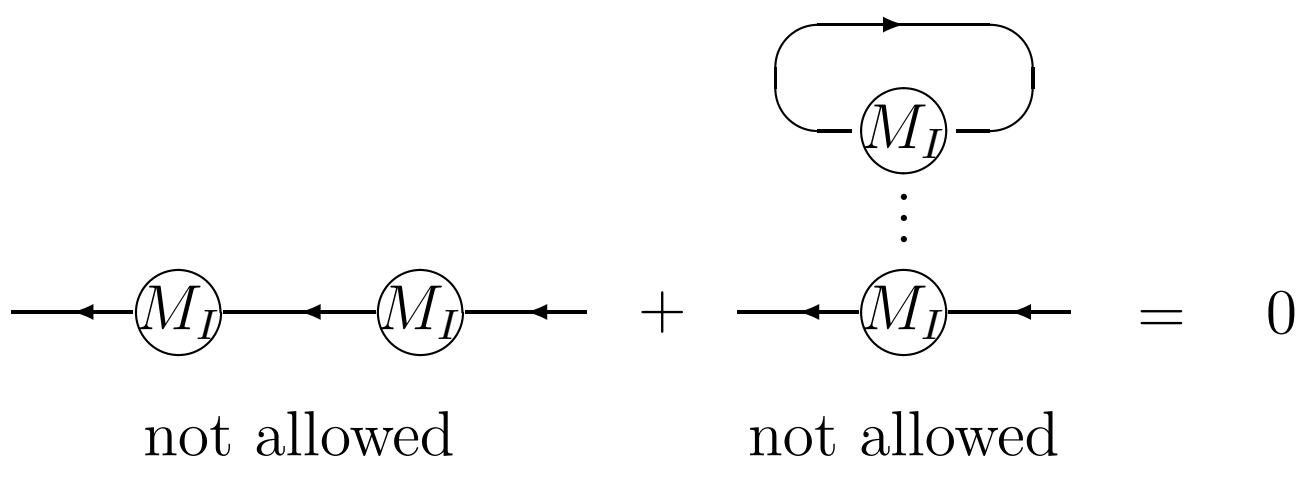

As in the case of $V_I$ both graphs are not allowed. Another example is

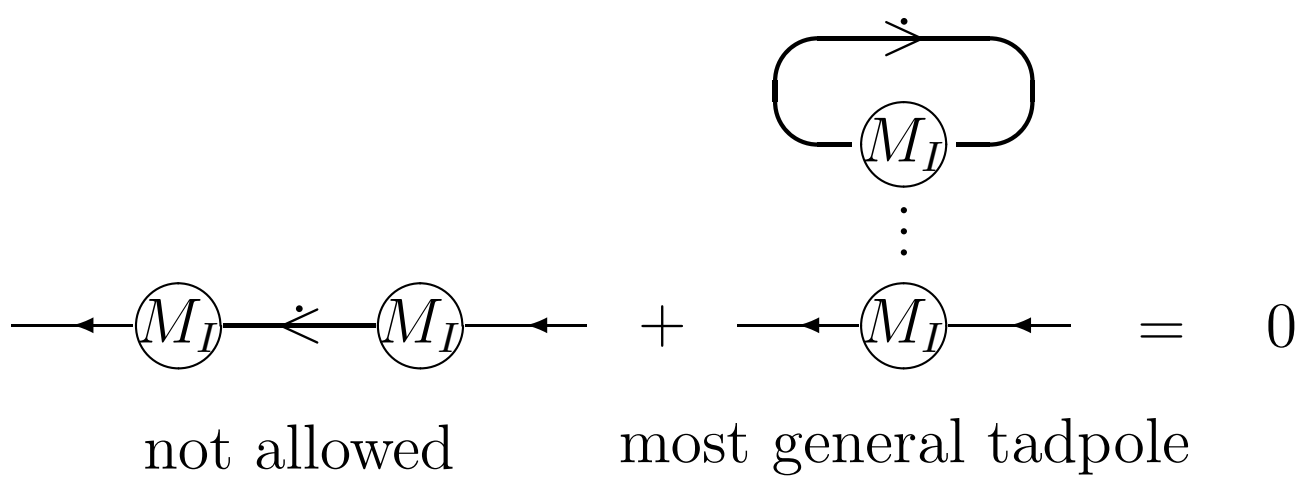

It is a general fact that all disallowed graphs can be paired with other disallowed graphs or with tadpoles and vice versa.

What we have to do is to "disallow" all tadpole graphs. Every $M_I$ can be surrounded by tadpoles which contribute with a universal multiplicative factor which can be absorbed in a redefinition of the instanton density $n_R$. Using this renormalized density $n_R$ the pairing is now perfect and the statement "every $M_I$ occurs only once" becomes true.

One further can show that in the presence of dynamical quarks this renormalized density has to be identified with the gluon condensate instead of the "bare" density because the same tadpoles contribute to the gluon condensate too.

### Selfconsistency Equation for the Quark Propagator

Quark loops are no longer possible because they cannot be connected to another part of the graph via a common instanton. All graphs which can contribute to the propagator are chains of different $M_I's$.

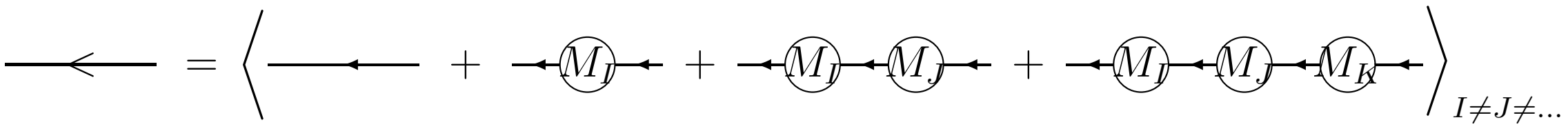

The $M_I's$ can be averaged independently

$$M(p) := i\langle M_I \rangle = \frac{n_R}{2\mu} p^2 \varphi'^2(p) \quad , \tag{56}$$

where $\varphi'$ is defined in appendix A. The resulting expression for the propagator now has the form

$$S = S_0 + S_0 \frac{M}{i} S_0 + S_0 \frac{M}{i} S_0 \frac{M}{i} S_0 + \ldots = (S_0^{-1} + M)^{-1} \tag{57}$$

where $M = M(p)$ is the momentum dependent mass defined above. There is just one thing to do: We have to solve the circular dependence but $\mu$ is just a number, which makes the solution very simple. Inserting (56) and (57) into (52) one gets the following equation for $\mu$

$$\mu = m + \int \bar{d}^4 p \, \frac{2\varphi'^2(p) M_p}{p^2 + (m + M_p)^2} (p^2 + m(m + M_p)) \tag{58}$$

which may be solved numerically for different current masses $m$.

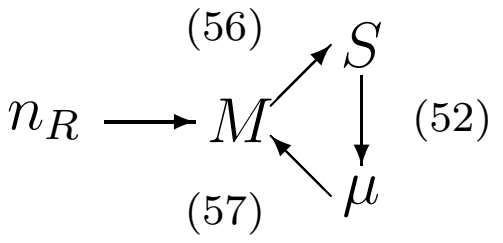

**Some Phenomenological Results**

In the chiral limit (58) reduces to

$$\mu^2 = n_R \int đ^4 p \, \frac{p^4 \varphi'^4(p)}{p^2 + M_p^2} = \alpha n_R \rho^2 + O(n_R^2) \tag{59}$$

$\mu^2$ is proportional to $n_R$ and thus $M$ is proportional to $\sqrt{n_R}$ in contrast to a linear dependence on $n_R$ obtained from a naive density expansion.

In the last expression the denominator has been expanded in the density and

$$\alpha = \rho^{-2} \int đ^4 p \, p^2 \varphi'^4(p) = 6.6 \tag{60}$$

is a universal number. For the standard values of $n_R$ and $\rho$ one gets

$$\begin{aligned} \mu_0^2 &= 6.6 n_R \rho^2 = (100\text{MeV})^2 \quad , \\ M(p=0) &= 7.7 \rho \sqrt{n_R} = 300\text{MeV} \quad . \end{aligned} \tag{61}$$

The exact solution of (59) which has been obtained numerically by iteration, differs from the leading density value by 15%:

$$M(0) = 345\text{MeV} \quad . \tag{62}$$

The momentum dependence of the quark mass is shown in figure 2. The mass $m + M(p)$ may be interpreted as the mass of a constituent quark. At high energies it tends to the current mass, at low momentum chiral symmety breaking occurs and the quark gets its constituent mass M(0). Note that this is not a pole mass but a virtual mass at zero momentum squared.

Let us now take into account a small current mass $m$ formally of the order $\sqrt{n_R}$. The selfconsistency equation now reads

$$1 = \frac{m}{\mu} + \frac{\mu_0^2}{\mu^2} + O(n_R) \quad . \tag{63}$$

Solving it for $\mu$ leads to

$$\mu_0 \quad \leq \quad \mu = \frac{1}{2}m + \sqrt{\frac{1}{4}m^2 + \mu_0^2} \quad \leq \quad m + \mu_0 \tag{64}$$

For the strange quark $\mu$ is increased by a factor of 2:

$$\mu(m_s = 150\text{MeV}) = 200\text{MeV} \tag{65}$$

It is interesting that $m_s + M(0)$ remains to be 300MeV. For zero momentum the increase of the current mass is just compensated by an equal decrease of the dynamical mass $M(0)$.

From the propagator one can obtain the quark condensate

$$\langle \bar{\psi}\psi \rangle := \lim_{x \to 0} \text{tr}_{CD}(S(x) - S_0(x)) = N_c \int đ^4 p \, \text{tr}_D(S(p) - S_0(p)) \quad . \tag{66}$$

In leading order in the density one gets

$$i\langle\bar{\psi}\psi\rangle = \frac{n_R N_c}{\mu} = \langle G^a_{\mu\nu} G^{\mu\nu}_a \rangle / 32\pi^2\mu \quad . \tag{67}$$

This leads to the following condensates for $u$,$d$ and $s$ quarks:

$$i\langle\bar{u}u\rangle = i\langle\bar{d}d\rangle = (250\text{MeV})^3 \quad , \quad \langle\bar{s}s\rangle = 0.5\langle\bar{u}u\rangle \quad . \tag{68}$$

For heavy quarks there exists a similar relation

$$i\langle\bar{\psi}\psi\rangle = \langle G^a_{\mu\nu} G^{\mu\nu}_a \rangle / 48\pi^2 m + O(m^{-3}) \quad , \tag{69}$$

which leads within 10% to the same value for the strange quark condensate. This nicely confirms the hypothesis that the strange quark can be treated as a light quark as well as a heavy quark. This hypothesis is used in heavy to light quark matching formulas.

**Large $N_c$ expansion**

Consider now the case of $N_f$ light quark flavors $u, d, s, \ldots$. The discussion of the one flavor case in the previous sections can be copied up to the pairing and cancelation of graphs which contain more than one $M_I$ (55). This is still true in the multiple flavor case but now both quark lines in (55) must have the same flavor because $M_I$ always connects quarks of the same flavor. So we have the theorem: "every $M_I$ occurs only once for each flavor". From this point on the discussion of the one flavor case breaks down because there are now graphs contributing to the propagator containing quark loops. The simplest new contribution has the form

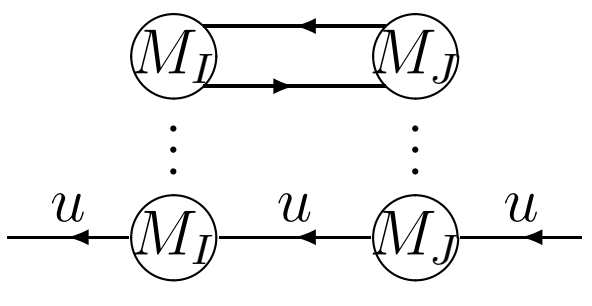

Is this contribution small in some sense ? Yes it is ! Quark loops are suppressed by a factor $1/N_c$. In perturbative context this is extensively discussed in [28], in instanton physics it was first used by [19]. Although 1/3 is not a very small number the large $N_c$ expansion seems to be a good approximation in various cases.

Consider a graph and add to it a new quark loop consisting of

| | | |
|---|---|---|
| $N$ | new instantons | $S \geq N$ . |
| $S$ | instanton scatterings | |

This multiplies the graph (see table 1) by a factor loop $= O(N_c^{1+N-S})$. The following cases are possible:

| Parameters | $n_R, \rho, N_c, (g_R)$ |
|---|---|
| Instanton density | $n = N/V_4 = n_R N_c \approx (200\text{MeV})^4$ |
| Instanton radius | $\rho \approx (600\text{MeV})^{-1}$ |
| Number of colors | $N_c = 3$ |
| Coupling constant | $g = g_R/N_c$ (not used in semicl. limit) |
| Gluon condensate | $\langle GG \rangle = \langle G^a_{\mu\nu} G^{\mu\nu}_a \rangle / 32\pi^2 =: n_R N_c$ |
| Quark condensate | $\langle \bar\psi\psi \rangle \approx 0.39\, N_c \rho^{-1} \sqrt{n_R} \approx (253\text{MeV})^3$ |
| Constituent mass | $M(p) \sim \rho\sqrt{n_R}$ |
| Quark mass | $M_{quark}(0) \approx 7.7\, \rho n_R^{1/2} \approx 300\text{MeV}$ |
| Gluon mass | $M_{gluon}(0) \approx 10.9\, \rho n_R^{1/2} \approx 420\text{MeV}$[24] |
| Ghost mass | $M_{ghost}(0) \approx 7.7\, \rho n_R^{1/2} \approx 300\text{MeV}$[24] |
| Meson correlator | $\langle \bar\psi\Gamma\psi(x)\bar\psi\Gamma\psi(0)\rangle_{trunc.} \sim N_c$ |
| Quark loop | 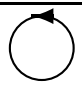 $\sim N_c$ |
| Instanton scattering | 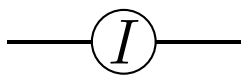 $\sim N_c^{-1} \rho n_R^{-1/2}$ |
| Instanton occurance | $\sim N_c n_R$ |
| Gluon loop | 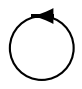 $\sim N_c^2 - 1$ |
| Instanton scattering | 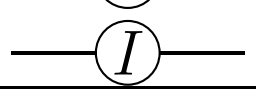 $\sim (N_c^2 - 1)^{-1}$ |

Table 1: Dependence of various quantities on the parameters of the instanton liquid model $n_R, \rho, N_c, (g_R)$.

$$\begin{array}{ll} \underline{1 + N - S} & \\ = 1 & \Longleftrightarrow M = N \Longleftrightarrow \text{all instantons are new} \\ & \Longleftrightarrow \text{the loop is disconnected} \\ = 0 & \Longleftrightarrow \text{one old instanton} \Longleftrightarrow \text{the loop is a tadpole} \\ < 0 & \text{the loop is supressed by at least one } 1/N_c \end{array}$$

Disconnected graphs are cancelled by the denominator and tadpoles have been absorbed in $n_R$. So quark loops are indeed suppressed in the limit $N_c \to \infty$. The same is true for gluon loops. This can be seen by the same argument using the $N_c$ dependences from table 1.

# 4 Four Point Functions

## Introduction

In the last section we have derived comprehensive Feynman rules within the zero mode aprroximation:

$$= \frac{1}{\not p + im} = S_0(p)$$

$p$ 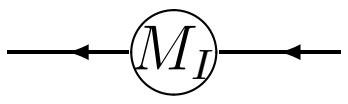 $q$

$$= \frac{1}{i\mu} \not p \psi_I(p) \psi_I^\dagger(q) \not q$$

A quark of a given flavor can scatter only once at a instanton $I$ via $M_I$. Tadpole graphs are not allowed, they are absorbed in the renormalized instanton density $n_R$ which has to be used when averaging over the instantons. In the large $N_c$ limit dynamical quark loops are supressed and $\mu$ can be determined by (58). In the chiral limit

$$\mu^2 = 6.6 n_R \rho^2 + O(n_R^2) \quad . \tag{70}$$

In this section we want to calculate 4 point functions in the case of two quark flavors of equal mass in the limit $N_c \to \infty$ within the zero mode approximation:

$$\bar{\delta}(p-s+r-q)\Pi_{\Gamma\Gamma'}(p,s,q,r) = \tag{71}$$

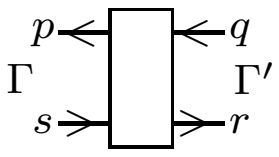

$$= -\int dxdydzdw\, e^{i(py-qz+rw-sx)} \langle 0|\mathcal{T}\bar{\psi}(x)\Gamma\psi(y)\bar{\psi}(z)\Gamma'\psi(w)|0\rangle \quad .$$

The $\psi$ fields are $u$ or $d$ quark fields arbitrarily mixed. Without restriction to generality we have taken the correlator to be a color singlet . These 4 point functions can be used to study meson correlators (see chapter 5)

$$\Pi(t) = $$

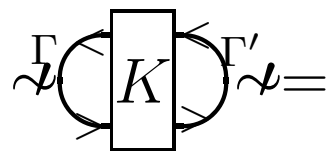

$$= \int(dpds)\int(dqdr)\Pi(p,s,q,r) = -\int dx\, e^{itx}\langle 0|\mathcal{T} j_\Gamma(x) j_{\Gamma'}(0)|0\rangle \quad ,$$

$$j_\Gamma(x) = \bar{\psi}\Gamma\psi(x) \quad , \quad j_{\Gamma'}(0) = \bar{\psi}\Gamma'\psi(0) \quad , \tag{72}$$

$$\int(dpds) = \int \bar{d}p\bar{d}s\,\bar{\delta}(p-s-t) \quad , \quad \int(dqdr) = \int \bar{d}q\bar{d}r\,\bar{\delta}(q-r-t) \quad ,$$

$$t = p - s = q - r \quad .$$

## Large $N_c$ approximation

The most general graph for the quark propagator is a sequence of different instantons $M_I$, according to the rules above. Similarly the most general graph for the triplet correlator $\langle(\bar{u}d)(\bar{d}u)\rangle$ consists of two quark propagators each containing every insanton only once. However, the $u$ and $d$ propagator may contain common instantons, e.g.

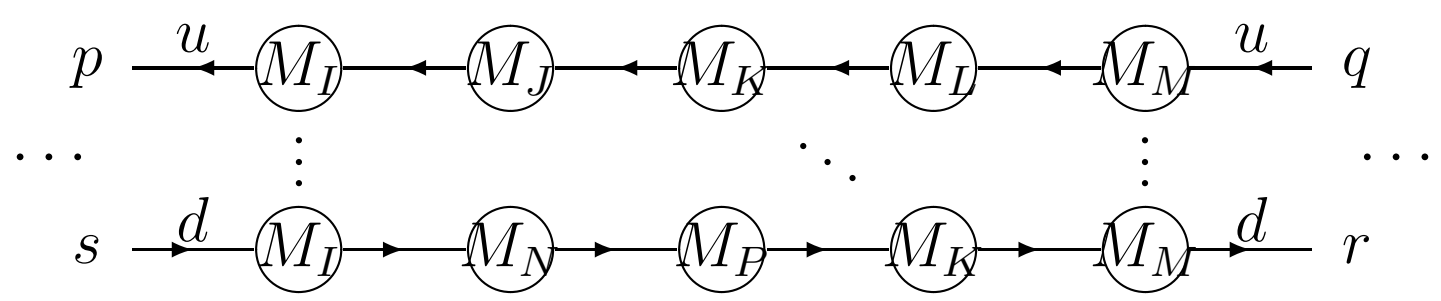

It is always assumed that the left and right hand sides of the graphs form color singlets. Non-common instantons can be averaged like in the propagator case to yield graphs of the form

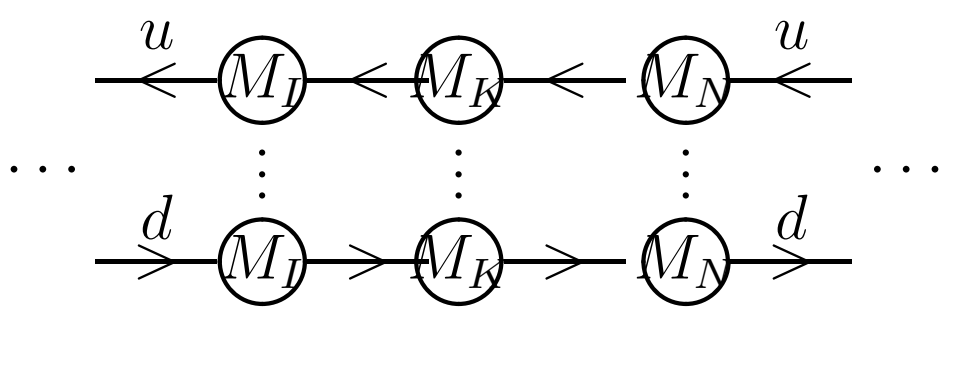

(73)

where the thick line represents the full propagator

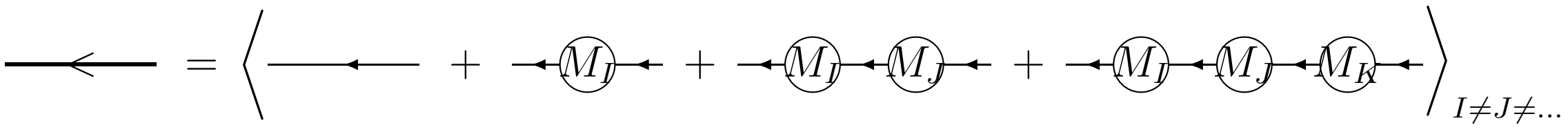

It can be shown that non planar diagrams like are suppressed by $1/N_c$, where again the left and right hand sides of the graphs have to form color singlets. So only the ladder diagrams shown in (73) contribute to the triplet correlator.

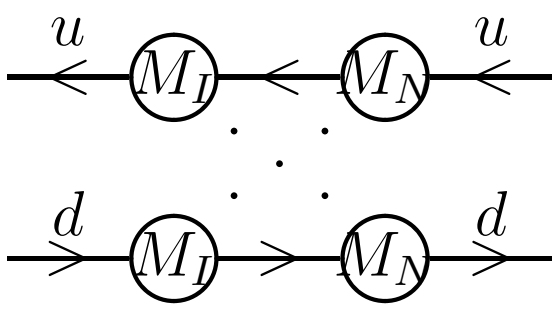

One might think that the mixed correlator $\langle(\bar{u}u)(\bar{d}d)\rangle$ is zero because the graphs necessarily include quark loops, or that only the two loop graphs contribute — but this is not the case! Let us first state the result and then discuss it. Graphs contributing to the mixed correlator are chains of quark bubbles

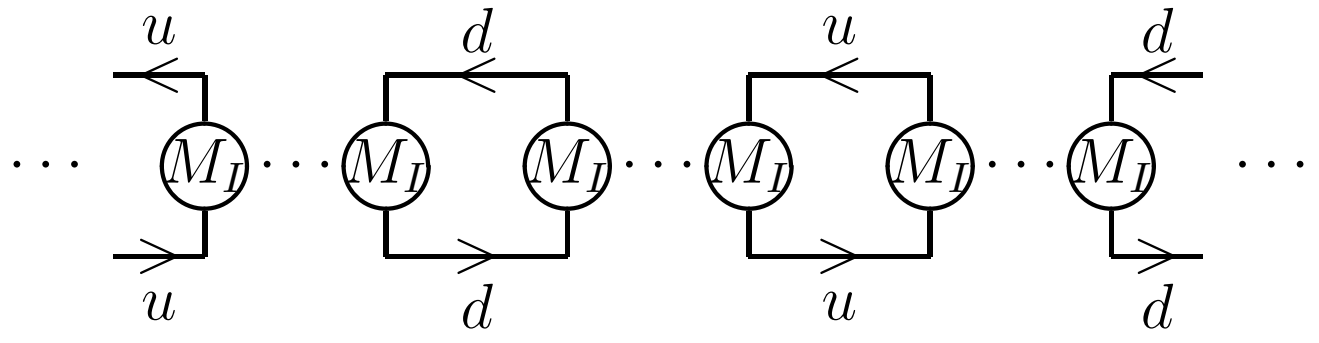

(74)

Application of the $N_c$ counting rules shows that this chain is of order $1/N_c$. Taking the color trace at the left and right hand side of the chain we see that the mixed correlator is of order $N_c$. Using (55) it is clear that the triplet correlator is of the same order.

What is wrong with the derivation of quark loop suppression in the last chapter ? The main assumption was that every graph containing a loop can be constructed from a graph not possessing this loop by simply adding the loop. Eliminating a loop from the bubble chain (74) yields a disconnected graph, but we only consider connected 4 point functions. So the quark loop chain cannot be constructed in a way needed to prove quark loop suppression.

In the case of meson correlators one can take another point of view. The disconnected two loop contribution is of order $N_c^2$ but (except for the scalar case) the contribution is zero. So the bubble chain is a subleading graph of order $N_c$ and nothing has been said about the form of subleading graphs.

Nevertheless, all connected graphs can be obtained starting from (74) by adding further instantons and bubbles — but now $N_c$ counting rules tell us that every attempt results in a $1/N_c$ suppression. Therefore the bubble chain is the most general leading order graph.

Nothing has to be changed for the correlator $\langle(\bar{d}d)(\bar{d}d)\rangle$ except that chain (74) must start with $d$.

To calculate the connected 4 point functions one must now average and sum up the chains. Alternatively this can be represented in recursive from usually called Bethe Salpeter equations:

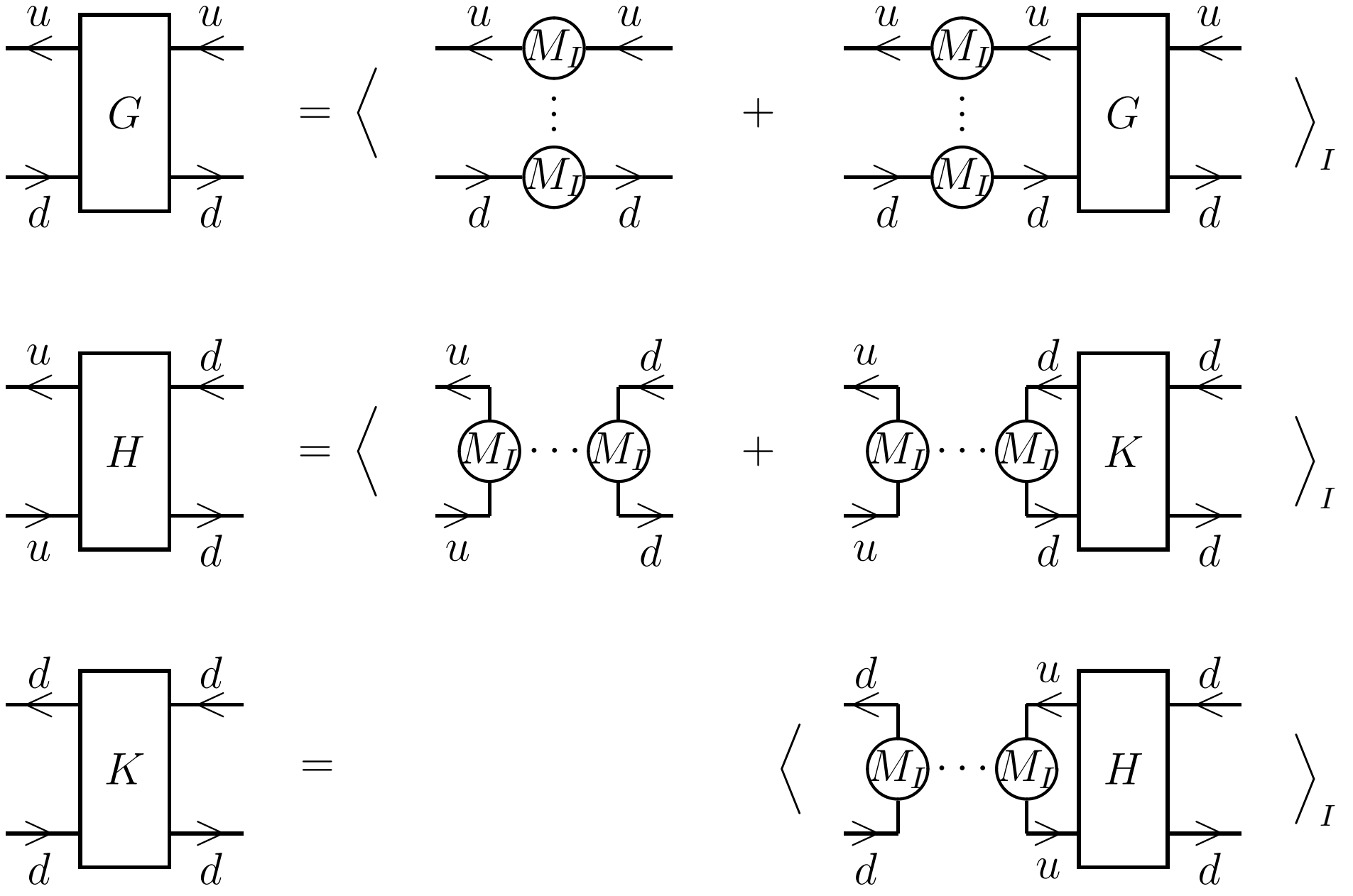

(75)

**Solution of the Bethe Salpeter Equations**

Before solving the BS equations we have to construct the kernel. The l.h.s. of the kernel always forms a color singlet because of the restriction to color singlet correlators.

Contracting the color and Dirac indices on the l.h.s. and using the formulas of appendix A one gets

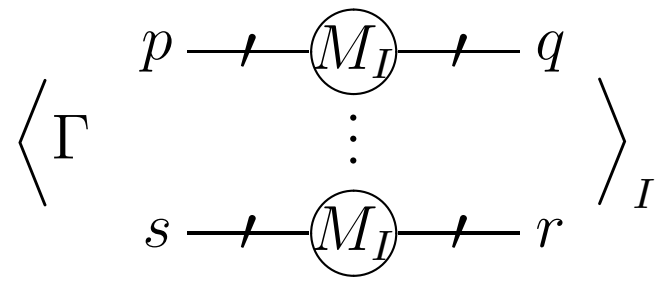

$$= \frac{1}{(i\mu)^2}\langle \not{r}\psi_I(r)\psi_I^\dagger(s)\not{s}\Gamma\not{p}\psi_I(p)\psi_I^\dagger(q)\not{q}\rangle_I = \qquad (76)$$

$$= -\frac{n_R}{\mu^2}p\varphi'(p)q\varphi'(q)r\varphi'(r)s\varphi'(s)\bar{\delta}(p-s+r-q)\left\langle \mathrm{tr}_D\left(\Gamma\frac{1\pm\gamma_5}{2}\right)\frac{1\pm\gamma_5}{2}\right\rangle_\pm \quad .$$

The kernel can now be determined to be

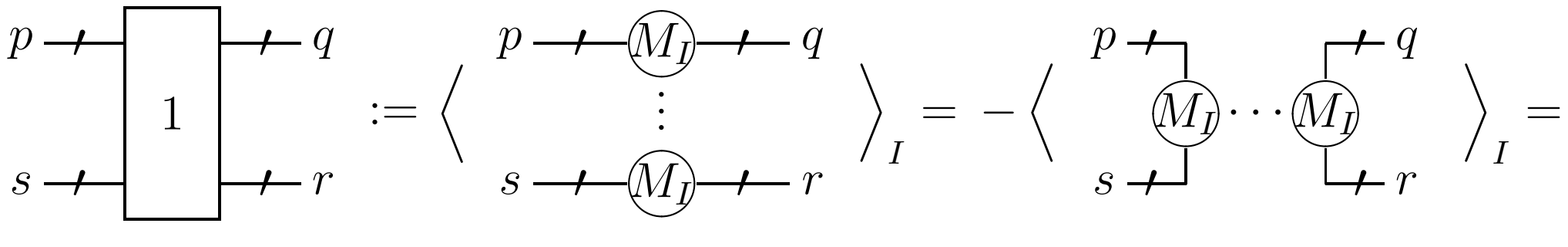

$$= -\frac{1}{n_R N_c}\sqrt{M_pM_qM_rM_s}\bar{\delta}(p-s+r-q)(\delta^{i_p}_{\ i_s}\delta^{i_r}_{\ i_q}+\gamma_5{}^{i_p}_{\ i_s}\gamma_5{}^{i_r}_{\ i_q})\delta^{\alpha_p}_{\alpha_s}\delta^{\alpha_r}_{\alpha_q} \quad . \qquad (77)$$

The result is just proportional to the nonlocal version of the 't Hooft vertex between color singlet states.

The solutions of the BS equations have a very similar structure:

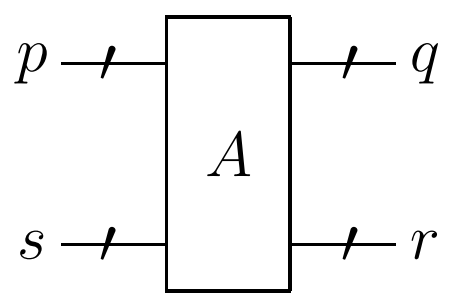

$$
\begin{aligned}
&:= -\frac{1}{n_R N_c}\sqrt{M_p M_q M_r M_s}\tilde{\delta}(p-s+r-q) \\
&\quad (A_0(t)\delta^{i_p}_{\ i_s}\delta^{i_r}_{\ i_q} + A_5(t){\gamma_5}^{i_p}_{\ i_s}{\gamma_5}^{i_r}_{\ i_q})\delta^{\alpha_p}_{\alpha_s}\delta^{\alpha_r}_{\alpha_q} \quad .
\end{aligned} \tag{78}
$$

$A_0$ and $A_5$ are scalar functions depending only on $t = p - s = q - r$. The proof is simple: The kernel has the structure (78) with $A_0 = A_5 = 1$. The product of two vertices yields the sames structure:

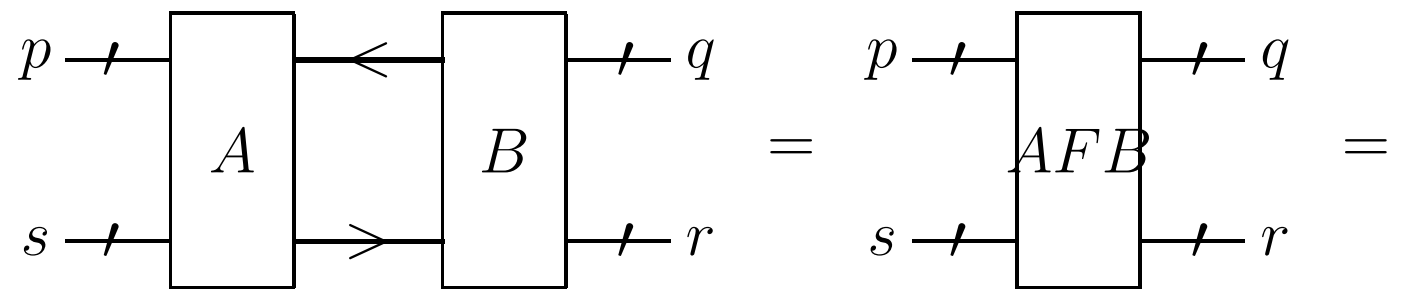

$$
\begin{aligned}
&= -\frac{1}{n_R N_c}\sqrt{M_p M_q M_r M_s}\tilde{\delta}(p-s+r-q) \\
&\quad (A_0 F_0 B_0(t)\delta^{i_p}_{\ i_s}\delta^{i_r}_{\ i_q} + A_5 F_5 B_5(t){\gamma_5}^{i_p}_{\ i_s}{\gamma_5}^{i_r}_{\ i_q})\delta^{\alpha_p}_{\alpha_s}\delta^{\alpha_r}_{\alpha_q} \quad ,
\end{aligned} \tag{79}
$$

$$
\begin{aligned}
F_0(t) &= -\int (dpds)\frac{1}{n_R} M_p M_s \mathrm{tr}_D(S(p)S(s)) \quad , \\
F_5(t) &= -\int (dpds)\frac{1}{n_R} M_p M_s \mathrm{tr}_D(S(p)\gamma_5 S(s)\gamma_5) \quad .
\end{aligned} \tag{80}
$$

In other words, the vertices of structure (78) build a closed algebra. The reason for this simple result is that the kernel is a simple product function up to the momentum conserving $\delta$.

Using (77), (78) and (79) the BS equations (75) reduce to primitive algebraic equations for $G_{0/5}(t)$, $H_{0/5}(t)$ and $K_{0/5}(t)$:

$$
\begin{aligned}
G_{0/5}(t) &= 1 + F_{0/5}(t)G_{0/5}(t) \quad , \\
H_{0/5}(t) &= -1 - F_{0/5}(t)K_{0/5}(t) \quad , \\
K_{0/5}(t) &= -F_{0/5}(t)H_{0/5}(t) \quad ,
\end{aligned}
$$

with the solution

$$
G = \frac{1}{1-F} \quad , \quad H = -\frac{1}{1-F^2} \quad , \quad K = \frac{F}{1-F^2} \tag{81}
$$

where we have suppressed the index 0/5 and the argument $t$.

**Triplet and Singlet Correlators**

Because of isospin symmetry $SU(2)_f$, mesons form triplets and singlets. Replacing $\bar{\psi}\psi$ in (72) by the triplet and singlet combinations (borrowing the notation from the pseudoscalar correlator)

$$\pi^0 = \frac{1}{\sqrt{2}}(\bar{u}u - \bar{d}d), \quad \pi^+ = \bar{u}d, \quad \pi^- = \bar{d}u, \quad \eta = \frac{1}{\sqrt{2}}(\bar{u}u + \bar{d}d), \tag{82}$$

one gets

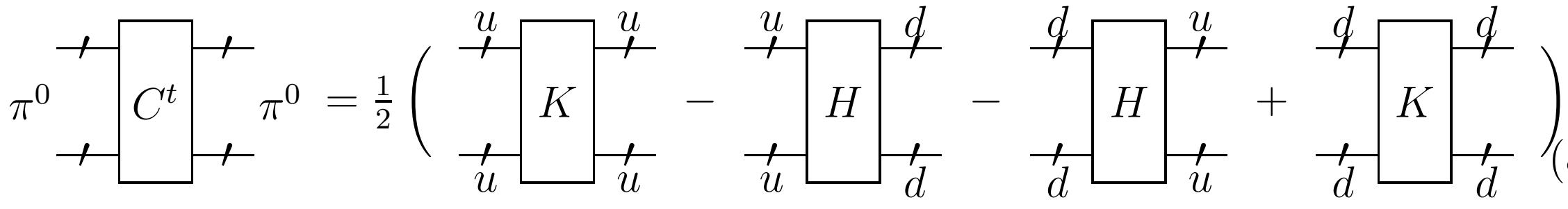

(83)

Therefore $C^t = K - H = \frac{1}{1-F}$. This coincides with $G = \frac{1}{1-F}$ for the charged triplet correlator $\langle(\pi^\pm)(\pi^\pm)\rangle$ as it should be. In the singlet case we get $C^s = K + H = -\frac{1}{1+F}$. When adding propagators in (78) to the external legs the final result for the connected 4 point function is

$$\Pi^{conn}_{\Gamma\Gamma'}(p,s,q,r) = -\frac{N_c}{n_R}\sqrt{M_p M_q M_r M_s} \;\; [C_0(t)\mathrm{tr}_D(S(s)\Gamma S(p))\mathrm{tr}_D(S(q)\Gamma S(r)) + C_5(t)\mathrm{tr}_D(S(s)\Gamma S(p)\gamma_5)\mathrm{tr}_D(S(q)\Gamma S(r)\gamma_5)] \tag{84}$$

$$C^{s/t}_{0/5}(t) = -\frac{1}{F_{0/5}(t) \pm 1} \quad \begin{array}{l} + \text{ for the singlet correlator} \\ - \text{ for the triplet correlator} \end{array} \tag{85}$$

$F_{0/5}(t)$ are defined in (80). The correlators of a singlet with a triplet current are zero as expected.

The following graphs may contribute to the disconnected part:

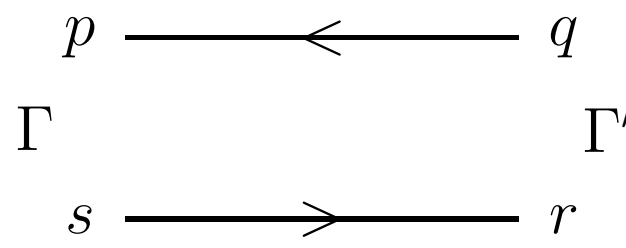

$$= N_c \mathrm{tr}_D(\Gamma S(p)\Gamma'(s))\bar{\delta}(p-q)\bar{\delta}(r-s)$$

(86)

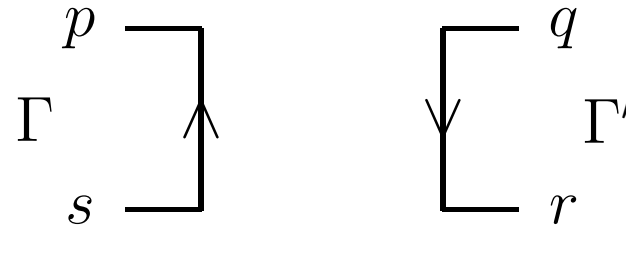

$$= -N_c^2 \mathrm{tr}_D(\Gamma S(p))\mathrm{tr}_D(\Gamma'(q))\bar{\delta}(p-s)\bar{\delta}(q-r)$$

depending on the flavor structure of the correlator. Note that the second two loop term is of the order $N_c^2$. However, as discussed above, in most applications it drops out or yields an uninteresting constant or only the connected part is considered anyway.

For the triplet and singlet case one gets

$$\bar{\delta}(p-s+q-r)\Pi^{disc}_{\Gamma\Gamma'}(psqr) = N_c \mathrm{tr}_D(\Gamma S(p)\Gamma' S(s))\bar{\delta}(p-q)\bar{\delta}(r-s) + \left\{ \begin{array}{ll} 0 & \text{for triplet} \\ 2\cdot(86) & \text{for singlet} \end{array} \right. \tag{87}$$

The 4 point functions obtained in this section will be discussed in the following chapters.

# 5 Correlators of Light Mesons

### Analytical Expressions

In the last chapter we have calculated various quark 4 point functions (71). The meson correlators or polarisation functions are just local versions of these vertices and can be obtained by simply setting $x = y$ and $z = w$. In momentum space the meson correlators have the form

$$\Pi_{\Gamma\Gamma'}(t) = \Pi^{disc}(t) + \Pi^{conn}(t) = $$ 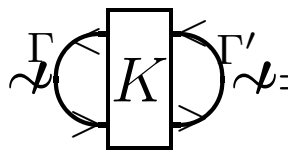 $$= \qquad (88)$$

$$= \int(dpds)\int(dqdr)\Pi(p,s,q,r) = -\int dx\, e^{itx}\langle 0|\mathcal{T} j_\Gamma(x) j_{\Gamma'}(0)|0\rangle \quad ,$$

$$j_\Gamma(x)^{s/t} = \frac{1}{\sqrt{2}}(\bar{u}\Gamma u(x) - \bar{d}\Gamma d(x)) \quad , \quad j_{\Gamma'}(0)^{s/t} = \frac{1}{\sqrt{2}}(\bar{u}\Gamma' u(0) - \bar{d}\Gamma' d(0)) \quad ,$$

$$\int(dpds) = \int đp\,đs\,\delta\!\!\!{}^{-}(p-s-t) \quad , \quad \int(dqdr) = \int đq\,đr\,\delta\!\!\!{}^{-}(q-r-t) \quad ,$$

$$t = p - s = q - r \quad .$$

From the explicit expressions of the 4 point functions obtained in the last chapter one can get, up to integration, analytical expressions for the meson correlators. The following list is a complete summary of all formulas needed to evaluate the meson correlators:

$$\begin{aligned}
\Pi^{disc}_{\Gamma\Gamma'}(t) &= N_c \int(dpds)\,\mathrm{tr}_D(\Gamma S(p)\Gamma' S(s)) \quad , \\
\Pi^{conn}_{\Gamma\Gamma'}(t) &= -N_c(C_0(t)\Gamma^0_\Gamma(t)\Gamma^0_{\Gamma'}(t) + C_5(t)\Gamma^5_\Gamma(t)\Gamma^5_{\Gamma'}(t)) \quad , \qquad (89) \\
C_{0/5}(t) &= -\frac{1}{F_{0/5}(t) \pm 1} \quad \begin{matrix} + \text{ for singlet correlator} \\ - \text{ for triplet correlator} \end{matrix} \quad , \\
\Gamma^0_\Gamma(t) &= \frac{1}{\sqrt{n_R}}\int(dpds)\,\sqrt{M_p M_s}\,\mathrm{tr}_D(S(p)\Gamma S(s)) \quad , \\
\Gamma^5_\Gamma(t) &= \frac{1}{\sqrt{n_R}}\int(dpds)\,\sqrt{M_p M_s}\,\mathrm{tr}_D(S(p)\Gamma S(s)\gamma_5) \quad , \\
F_0(t) &= \frac{-1}{n_R}\int(dpds)\,M_p M_s \mathrm{tr}_D(S(p)S(s)) \quad , \\
F_5(t) &= \frac{-1}{n_R}\int(dpds)\,M_p M_s \mathrm{tr}_D(S(p)\gamma_5 S(s)\gamma_5) \quad , \\
S(p) &= \frac{1}{\not{p} + i(m + M_p)} \quad , \quad M_p = \frac{n_R}{2\mu}p^2\varphi'^2(p) \quad , \quad , \\
p\varphi'(p) &= 2\pi\rho z \frac{\partial}{\partial z}[I_0(z)K_0(z) - I_1(z)K_1(z)]_{z=p\rho/2} \quad , \\
\mu &= m + \int đ^4p\, \frac{2\varphi'^2(p)M_p}{p^2 + (m + M_p)^2}(p^2 + m(m + M_p)) \quad , \qquad (90) \\
n_R N_c &= (200\mathrm{MeV})^4 \quad , \quad \rho = (600\mathrm{MeV})^{-1} \quad .
\end{aligned}$$

**Analytical Results**

Performing the Dirac traces leads to the following expressions:

$$
\begin{aligned}
F_{0/5}(t) &= -\frac{4}{n_R}\int(dpds)\frac{M_pM_s(\pm(ps)-\tilde{M}_p\tilde{M}_s)}{(p^2+\tilde{M}_p^2)(s^2+\tilde{M}_s^2)} \quad , \\
\Gamma^{0/5}_{1/5}(t) &= \frac{4}{\sqrt{n_R}}\int(dpds)\frac{\sqrt{M_pM_s}(\pm(ps)-\tilde{M}_p\tilde{M}_s)}{(p^2+\tilde{M}_p^2)(s^2+\tilde{M}_s^2)} \quad , \\
\Gamma^{5}_{\mu 5}(t) &= \frac{4i}{\sqrt{n_R}}\int(dpds)\frac{\sqrt{M_pM_s}(\tilde{M}_ps_\mu-\tilde{M}_sp_\mu)}{(p^2+\tilde{M}_p^2)(s^2+\tilde{M}_s^2)} \quad , \\
\tilde{M}_p &= m+M_p
\end{aligned}
\tag{91}
$$

All other vertices $\Gamma$ are zero. Consider the one instanton vertex (77) (the kernel). It contributes only to the scalar and pseudoscalar correlator. From this observation one may have predicted that the connected part of all other channels is small because a contribution has to be a multi-instanton effect. Indeed, they are all zero as seen above except for the axial correlator. Due to an extra factor $M \sim \sqrt{n_R}$ in the numerator of $\Gamma^5_{\mu 5}$ the connected part of the axial correlator is suppressed by $O(n_R)$ therefore it is small as expected and will be neglected in the following.

Furthermore we will restrict ourself to the *chiral limit*, taking $m = 0$. Using the selfconsistency equation (90) one can see that

$$F_5(t=0) = 1 \tag{92}$$

is leading to a pole at $t = 0$ in the pseudoscalar triplet correlator due to the $F_5(t) - 1$ denominator in (89). This is the massless Goldstone pion one expects in the chiral limit. A more extensive discussion can be found in [19]. On the other side in the singlet correlator the minus sign is replaced by a positive sign and there is no Goldstone boson in this case. Thus, the (two flavor) $\eta'$ meson is massive ! Unfortunately we cannot make any reliable prediction of the $\eta'$ mass because the kernal is very repulsive in this channel and no boundstate is formed. There have to be other attractive forces, e.g. confinement forces, to built an $\eta'$ boundstate. Similar things happen in the scalar triplet channel (compare figure C and 8). But the most important thing is that there is no massless pseudoscalar singlet meson which is an important step towards discussing the $U(1)_A$-problem.

It is interesting to see that in leading order in the instanton density $F_0(t) = -F_5(t)$, which leads to a massless pole in the scalar singlet correlator. Numerically the $\sigma$-meson indeed turns out to be very light. The experimental situation is rather unclear.

**Spectral Representation**

To extract phenomenological information from the meson correlators we make use of the spectral representation

$$\Pi(p) = \int dx\, e^{ipx}\langle 0|\mathcal{T} j_\Gamma(x) j_{\Gamma'}(0)|0\rangle = \int_0^\infty d\sigma^2\, D(\sigma,x)\rho(\sigma^2) \tag{93}$$

where

$$\rho(p^2) = (2\pi)^3 \sum_n \delta(p - q_n) \langle 0 | j_\Gamma(0) | n \rangle \langle n | j'_\Gamma(0) | 0 \rangle \tag{94}$$

is the spectral density and

$$D(m, x) = \int đ^4 p \frac{e^{-ipx}}{p^2 + m^2} = \frac{1}{4\pi^2 x^2}(mx) K_1(mx) \tag{95}$$

is the free propagator of mass $m$ in coordinate representation. We have chosen the coordinate representation of $\Pi$ to be able to compare the plots directly with lattice calculations and with numerical studies of the instanton liquid [21].

The spectrum consists of mesonic resonances and the continuum contributions. If one is only interested in the properties of the first resonance one might approximate the rest of the spectrum by the perturbatively calculated continuum.

One might think that the disconnected part only contributes to the continuum and the connected part will yield the boundstates. But this is not the case. On one hand, Bethe Salpeter equations have bound as well as continuum solutions. On the other hand consider a theory with weak attraction between particles of mass m. It is clear that there is only a cut above $2m$ and no boundstate pole in the free loop. However in the exact polarization function only a small portion of the continuum will be used to form a pole just below the threshold because the attraction is only weak. The Euclidian correlator will hardly be changed. Therefore, assuming weak attraction, we can already estimate the boundstate mass from the disconnected part. Of course in this example we need not calculate anything because we know that the bounstate mass is approximately $2m$ with errors of the order of the strength of the interaction.

Assuming that all other forces neglected in QCD so far, especially perturbative corrections, are small and attractive in the vector and axialvector channel, we can obtain boundstate masses although in these channels up to our approximation there is no connected part. But things are less trivial than in the example above because the quarks do not posses a definite mass and we have to inspect the correlator to extract the meson masses.

Let us start with the scalar and pseudoscalar correlator. The lowest resonance of mass $m_*$ is coupled to the current with strength

$$\lambda_* = \langle 0 | j_{1/5}(0) | p \rangle \quad . \tag{96}$$

The rest of the spectrum is approximated by the continuum starting at the threshold $E_*$. * means $\pi$, $\eta$, $\delta$ or $\sigma$ (see table 2). $E_*$ is typically of the order 1.5 GeV and therefore the continuum can be calculated perturbatively. The spectrum thus has the form

$$\rho_{1/5}(s) = \lambda_*^2 \delta(s - m_*^2) + \frac{3s}{8\pi^2} \Theta(s - E_*^2) \quad . \tag{97}$$

Inserting $\rho$ into (93) one gets

$$\Pi_{1/5}^{fit}(x) = \lambda_*^2 D(m_*, x) + E(E_*, x) \quad , \tag{98}$$

| Correlator | | $\Gamma = \Gamma'$ | I=1 | I=0 |
|---|---|---|---|---|
| Pseudoscalar | $\Pi_5 = \langle j_5 j_5 \rangle$ | $i\gamma_5$ | $\pi$ | $\eta'$ |
| Scalar | $\Pi_1 = \langle j_1 j_1 \rangle$ | $1\!\!1_D$ | $\delta$ | $\sigma$ |
| Vector | $\Pi_{\mu\mu} = \langle j_\mu j_\mu \rangle$ | $\gamma_\mu$ | $\rho$ | $\omega$ |
| Axialvector | $\Pi^5_{\mu\mu} = \langle j^5_\mu j^5_\mu \rangle$ | $\gamma_\mu\gamma_5$ | $a_1$ | $f_1$ |

Table 2: Mesonic correlators

$$E(E_*, x) = \frac{3}{\pi^4 x^6} \frac{(E_* x)^3}{16} (2K_3(E_* x) + (E_* x) K_2(E_* x)) \overset{x\to 0}{\longrightarrow} \frac{3}{\pi^4 x^6} \quad . \tag{99}$$

In the next section $m_*$, $\lambda_*$ and $E_*$ are obtained by fitting the phenomenological ansatz $\Pi^{fit}(x)$ to the theoretical curve $\Pi^{sum}(x)$ in the Euclidian region where the theoretical calculation is reliable.

Consider now the vector and axial vector correlator. The vector current is conserved, thus the correlator is transverse and only the vector meson can contribute. In the chiral limit the same holds true for the axial current. In the singlet channel one has to be careful because there are two currents. A conserved one and a gauge invariant one which contains an anomaly. Up to now we have only calculated the correlator of the conserved current. Nevertheless to leading order in the instanton density the two correlators coincide and should be both conserved.

For conserved vector and axial currents the spectral function is transverse:

$$\rho^{(5)}_{\mu\nu}(p^2) = (-\delta_{\mu\nu} + \frac{p_\mu p_\nu}{p^2}) \rho^{(5)}_T(p^2) \quad . \tag{100}$$

The coupling of the vector and axial meson to the current is given by

$$i\lambda_* \epsilon_\mu = \langle 0 | j^{(5)}_\mu(0) | p \rangle \tag{101}$$

where $\epsilon_\mu$ is the meson polarization. The spectral and polarization functions have the form

$$\begin{aligned} -\rho^{(5)}_{\mu\mu}(s) &= 3\lambda_*^2 \delta(s - m_*^2) + \frac{3s}{4\pi^2} \Theta(s - E_*^2) \quad , \\ -\Pi^{fit(5)}_{\mu\mu}(x) &= 3\lambda_*^2 D(m_*, x) + 2E(x) \quad . \end{aligned} \tag{102}$$

Here * means $\rho$, $\omega$, $a_1$ or $f_1$ (see table 2).

### Plot & Fit of Meson Correlators

The meson correlators are shown in figure 3 - 8. The numerical evaluation of the integrals is discussed in appendix B. The correlators are normalized to the free correlator

$$\Pi^0_{1/5}(x) = \frac{3}{\pi^4 x^6} \quad , \quad \Pi^{0(5)}_{\mu\mu} = -\frac{6}{\pi^4 x^6} \quad . \tag{103}$$

The diagrams therefore show the deviation from the perturbative behaviour. The meson parameters obtained by fitting the parameter ansatz to the theoretical curve are summarazied in table 3. Shuryak & Verbarshot [21] have obtained the same parameters from a numerical investigation of the instanton liquid model. Their values are also shown in table 3. The parameters of the vector channel coincide extremely well with ours. For the $\pi$ and $\sigma$ meson there is a large discrepancy in the masses but this is not surprising: We are working in the chiral limit thus the pion mass has to be zero. A similar argument holds for the $\sigma$ meson as discussed above. The couplings fit very well. The discrepancy in the axial channels can have various origins which are under investigation. Alternatively one may directly compare the graphs. They coincide very well even in cases where a spectral fit does not work very well like in the $\delta$ and $\eta'$ channel. The conclusion is that the terms neglected in our analytical treatment, but included in the numerical study [21], are small and usually give an correction less than 10%. These are contributions from nonzero modes and higher order corrections in $1/N_c$. This is again an example for the surprisingly high accuracy of the $1/N_c$ expansion. In the case of strange quarks the nonzero mode contributions will become more important.

Finally, one should compare the numbers with experiment. As far as known, these numbers are also listed in table 3. A general discussion of the meson correlators and comparision with experimental results can be found in [21].

# 6 Conclusions

Various QCD correlators have been calculated in the instanton liquid model in zeromode approximation and $1/N_c$ expansion. The $1/N_c$ expansion should be seen as a substitute for the density expansion which fails in the presence of light quarks. We have extended the work [19] by including dynamical quark loops. In contrast to the original "perturbative" $1/N_c$ expansion [28], not all quark loops are suppressed. In the flavor singlet meson correlators a chain of quark loops survives the $1/N_c$ limit, causing the absence of a Goldstone boson in the pseudoscalar singlet channel. The analytical expressions for the meson correlators have been evaluated numerically and meson masses and couplings have been obtained from a spectral fit. Except for the pseudoscalar singlet and the scalar triplet channel, the spectral ansatz matches very well the theoretical curve. The numbers obtained in this way are consistent with those obtained by numerical studies of the instanton liquid [21] within 10%. Comparision with experiment is also quite satisfactory.

**Acknowledgements**

I want to thank the "Deutsche Forschungsgemeinschaft-Gemeinschaft" for supporting this work.

# A Zeromode Formulas

The covariant derivative $\not{D}$ in the background of an instanton has one zeromode

$$(i\not{\partial} - \not{A}_I)\psi_I = 0 \tag{104}$$

| Meson | $I^G(J^{PC})$ | $m_*$[MeV] | $\sqrt{\lambda_*}$[MeV] | $E_*$[MeV] | source |
|---|---|---|---|---|---|
| $\pi$ | $1^-(0^{-+})$ | 0 | 508±1 | 1276±33 | $1/N_c$ |
| | | 142±14 | 510±20 | 1360±100 | simulation |
| | | 138 | 480 | — | experiment |
| $\eta'$ | $0^+(0^{-+})$ | $\neq 0$ | ? | ? | $1/N_c$ |
| | | $\neq 0$ | ? | ? | simulation |
| | | 960 | ? | — | experiment |
| $\delta$ | $1^-(0^{++})$ | $\neq 0$ | ? | ? | $1/N_c$ |
| | | $\neq 0$ | ? | ? | simulation |
| | | 970 | ? | — | experiment |
| $\sigma$ | $0^+(0^{++})$ | 433±3 | 506±3 | 1446±20 | $1/N_c$ |
| | | 543 | 500 | 1160 | simulation |
| | | ? | ? | — | experiment |
| $\rho$ | $1^+(1^{--})$ | 930±5 | 408±4 | 1455±33 | $1/N_c$ |
| | | 950±100 | 390±20 | 1500±100 | simulation |
| | | 780 | 409±5 | — | experiment |
| $\omega$ | $0^-(1^{--})$ | 930±5 | 408±4 | 1455±33 | $1/N_c$ |
| | | ? | ? | ? | simulation |
| | | 780 | 390±5 | — | experiment |
| $a_1$ | $1^-(1^{++})$ | 1350±200 | 370±30 | 1050±80 | $1/N_c$ |
| | | 1132±50 | 305±20 | 1100±50 | simulation |
| | | 1260 | 400 | — | experiment |
| $f_1$ | $0^+(1^{++})$ | 1350±200 | 370±30 | 1050±80 | $1/N_c$ |
| | | 1210±50 | 293±20 | 1200±50 | simulation |
| | | 1285 | ? | — | experiment |

Table 3: Meson mass $m_*$, coupling constant $\lambda_*$ and continuum threshold $E_*$ obtained within the instanton liquid model in this work ($1/N_c$ expansion), from numerical simulation and from experiment.

The solution for an instanton of topological charge $Q_I = \pm 1$ in singular gauge in coordinate and momentum space is given in terms of modified Bessel functions $I_{0/1}$ and $K_{0/1}$

$$\begin{aligned}
\psi_I(x+z_I) &= \sqrt{2}\varphi(x)\not{x}\chi_I^{\pm} \quad , \quad \varphi(x) = \frac{\rho}{\pi|x|(x^2+\rho^2)^{3/2}} \\
\psi_I(p) &= \frac{\sqrt{2}\varphi'(p)}{|p|}e^{ipz_I}\chi_I^{\pm} \quad , \quad \varphi(p) = \int d^4x\, e^{ipx}\varphi(x) \\
\varphi'(p) &= \pi\rho^2\frac{d}{dz}[I_0(z)K_0(z) - I_1(z)K_1(z)]_{z=|p|\rho/2} = \begin{cases} -\frac{2\pi\rho}{|p|} & : \ p\rho \ll 1 \\ -\frac{12\pi}{p^4\rho^2} & : \ p\rho \gg 1 \end{cases}
\end{aligned} \tag{105}$$

$$\rho^2\int d^4x\,\varphi^2(x) = \int d^4x\,\varphi^2(x)x^2 = \int d\!\!\!^-{}^4p\,\varphi'^2(p) = \frac{1}{2}$$

$$\chi_I^{\pm} \sim \gamma_\mu \frac{1\mp\gamma_5}{2}\tau_\mu^{\mp}\Big|_{\textit{some color \& Dirac column}} \quad , \quad \tau_\mu^{\mp} = (\underline{\tau}, \pm i)$$

As usual only the projectors can be given in a covariant way[4]:

$$\chi_I^{\pm}\bar{\chi}_J^{\pm} = \frac{1}{16}(\gamma_\mu\gamma_\nu\frac{1\pm\gamma_5}{2})(U_I\tau_\mu^{\mp}\tau_\nu^{\pm}U_J^{\dagger}) \tag{106}$$

$$\chi_I^{\pm}\bar{\chi}_J^{\mp} = \mp\frac{i}{4}(\gamma_\mu\frac{1\mp\gamma_5}{2})(U_I\tau_\mu^{\mp}U_J^{\dagger}) \tag{107}$$

$U_{I/J} \in SU(N_c)$ is the orientation matrix of the instanton $I/J$. Although multi-instanton effects are studied in this work formulas like (106) concerning the overlap of different instantons are not needed. (106) is only needed for $I = J$ in the following special cases:

$$N_C\langle\chi_I^{\pm}\bar{\chi}_I^{\pm}\rangle_I = \mathrm{tr}_D\chi_I^{\pm}\bar{\chi}_I^{\pm} = \frac{1}{2}(\frac{1\pm\gamma_5}{2}) \tag{108}$$

$$\bar{\chi}\chi = \mathrm{tr}_{CD}\chi\bar{\chi} = 1 \tag{109}$$

Taking space, color and Dirac part together we get

$$\psi_I^{\dagger}(p)\psi_I(p) = 2\varphi'^2(p) \quad \Longrightarrow \quad \int d\!\!\!^-{}^4p\,\psi_I^{\dagger}(p)\psi_I(p) = 1 \tag{110}$$

# B Numerical Evaluation of Integrals

The integral expressions for the meson correlators have been evaluated numerically. Two types of operations have to be performed:

1. Convolution of Lorentz covariant functions $(F_{0/5}, \Gamma_\Gamma)$
2. Fourier transformation (FT) of the correlators to coordinate space

[4]In Euclidian space the bar operation is the same as the hermitian conjugate: no $\gamma_0$ is introduced, thus $\bar{\chi} = \chi^{\dagger}$.

**Fourier Transformation**

Let us first consider the FT generalized to $d$ dimensions:

$$\hat{f}_{\mu_1\ldots\mu_n}(x) = F_d\{f_{\mu_1\ldots\mu_n}\}(x) = \int d\!\!\bar{}^d p\, e^{-ipx} f_{\mu_1\ldots\mu_n}(p) \tag{111}$$

For a scalar spherically symmetric function $f = f(|p|)$ the FT reduces to a one dimensional integral

$$F_d\{f(|p|)\}(x) = \int_0^\infty \left(\frac{m}{2\pi|x|}\right)^{d/2} f(m) J_{d/2-1}(m|x|)|x|\, dm \tag{112}$$

where $J_\nu$ are Bessel functions.

If $x$ is not too large and $f$ decays rapidly, the integration can be performed with gaussian (or other) integration methods. If the decay is too slow one has to subtract the asymptotic part from $f$ thus improving convergency. The FT of the asymptotic part can be performed analytically and has to be added to the numerical FT of the reduced function.

The FT of a general Lorentz covariant function can also be reduced to (112) with (formally) an increased dimension $d$:

$$\begin{aligned} F_d\{p_\mu f(|p|)\}(x) &= 2\pi i x_\mu F_{d+2}\{f\}(x) \qquad (113)\\ F_d\{p_\mu p_\nu f(|p|)\}(x) &= \frac{1}{d-1}\Big[(\delta_{\mu\nu} - \frac{x_\mu x_\nu}{x^2})F_d\{p^2 f(|p|)\}(x) - \\ &\quad (\delta_{\mu\nu} - d\frac{x_\mu x_\nu}{x^2})(4\pi F_{d+2}\{f\}(x) - 4\pi^2 x^2 F_{d+4}\{f\}(x))\Big] \\ &\ldots \end{aligned}$$

**Convolution**

Now we treat the convolution integrals

$$f * g(p) = \int d\!\!\bar{}^d q\, f(q)\cdot g(p-q) \tag{114}$$

This type of integral can be reduced to the FT discussed above:

$$f * g(p) = F_d^{-1}\{F_d\{f\}\cdot F_d\{g\}\} \tag{115}$$

This is a quick and easy method for evaluating convolution integrals. For the disconnected part of the correlators it has the advantage that in coordinate representation $F_d^{-1}$ can be dropped. The disadvantage of this formula is that the FTs involve oscillating integrals which are numerically problematic. If the back-transformation $F_d^{-1}$ is needed as in the case of $F_{0/5}$ and $\Gamma_\Gamma$ it is better to perform the convolution directly. Similar to the FT the convolution can be reduced to the scalar case. The convolution of two scalar functions can further be reduced to a two dimensional integral:

$$f * g(p) = \frac{(d/2-1)!}{2\pi^{d/2+1}(d-2)!}\int_0^\infty dr \int_0^\pi d\theta\, f(r) g(\sqrt{p^2 - 2|p|r\cos\theta + r^2})(r\sin\theta)^{d-2} r \tag{116}$$

It is again evaluated with gaussian integration methods. The second advantage is that there are no problems with slowly decaying functions. Sometimes there are large cancelations between different terms. In this case it is essential to use nonadaptive integration methods because they will not result in a loss of accuracy.

The explicit reduction of the various correlators to the basic forms (112) and (116) is more or less trivial. The selfconsistency equation has been solved by iteration. The results are plotted in figure 3 - 8.

# C Figures

ifig13.pic

Figure 1: $S$ is minimal at the river, small in the valley and large in the mountains, Therefore $Z = \int dx_1 dx_2 \, e^{-S[x_1,x_2]}$ is dominated by the valley.

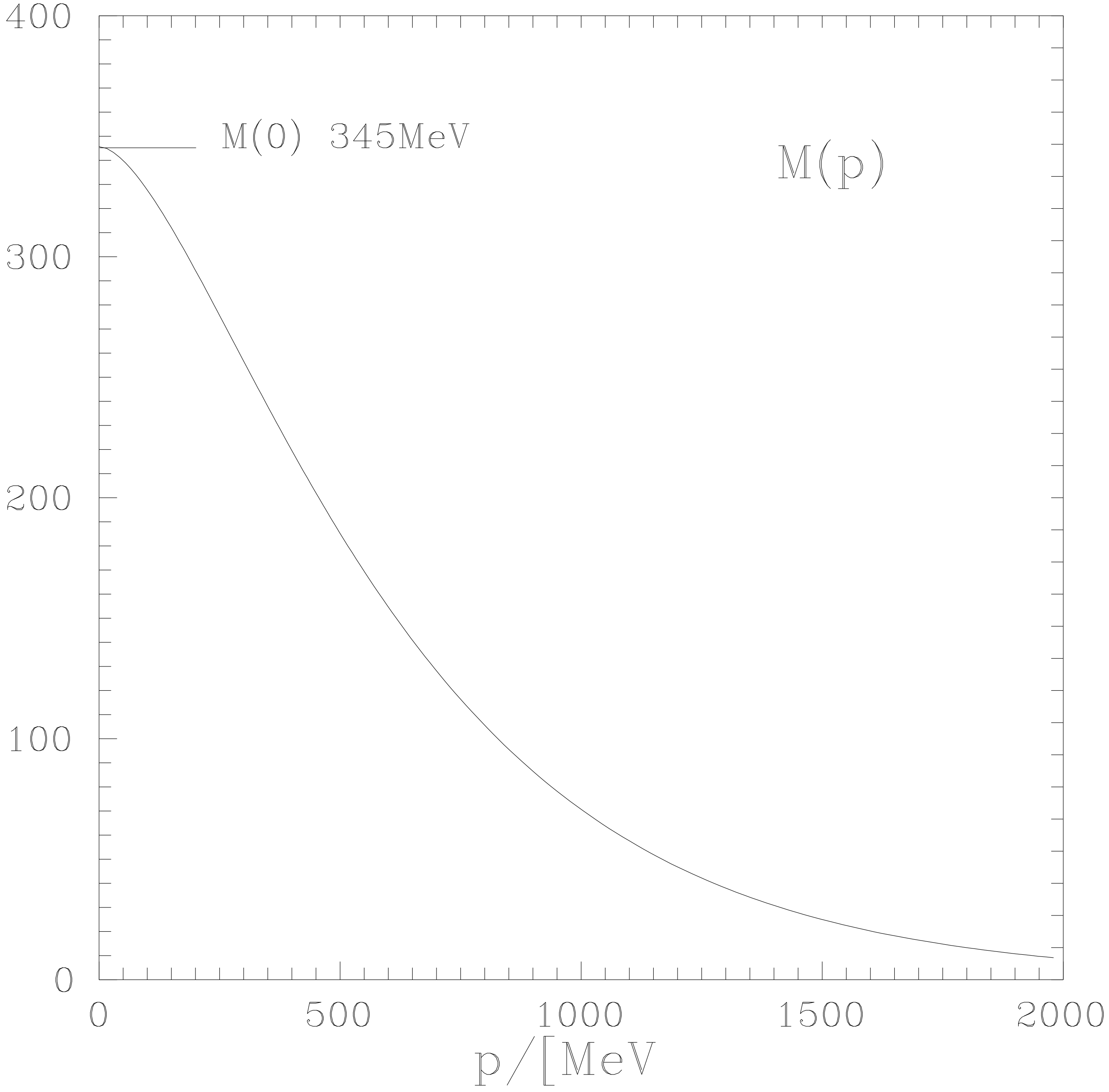

Figure 2: Constituent quark mass M(p).

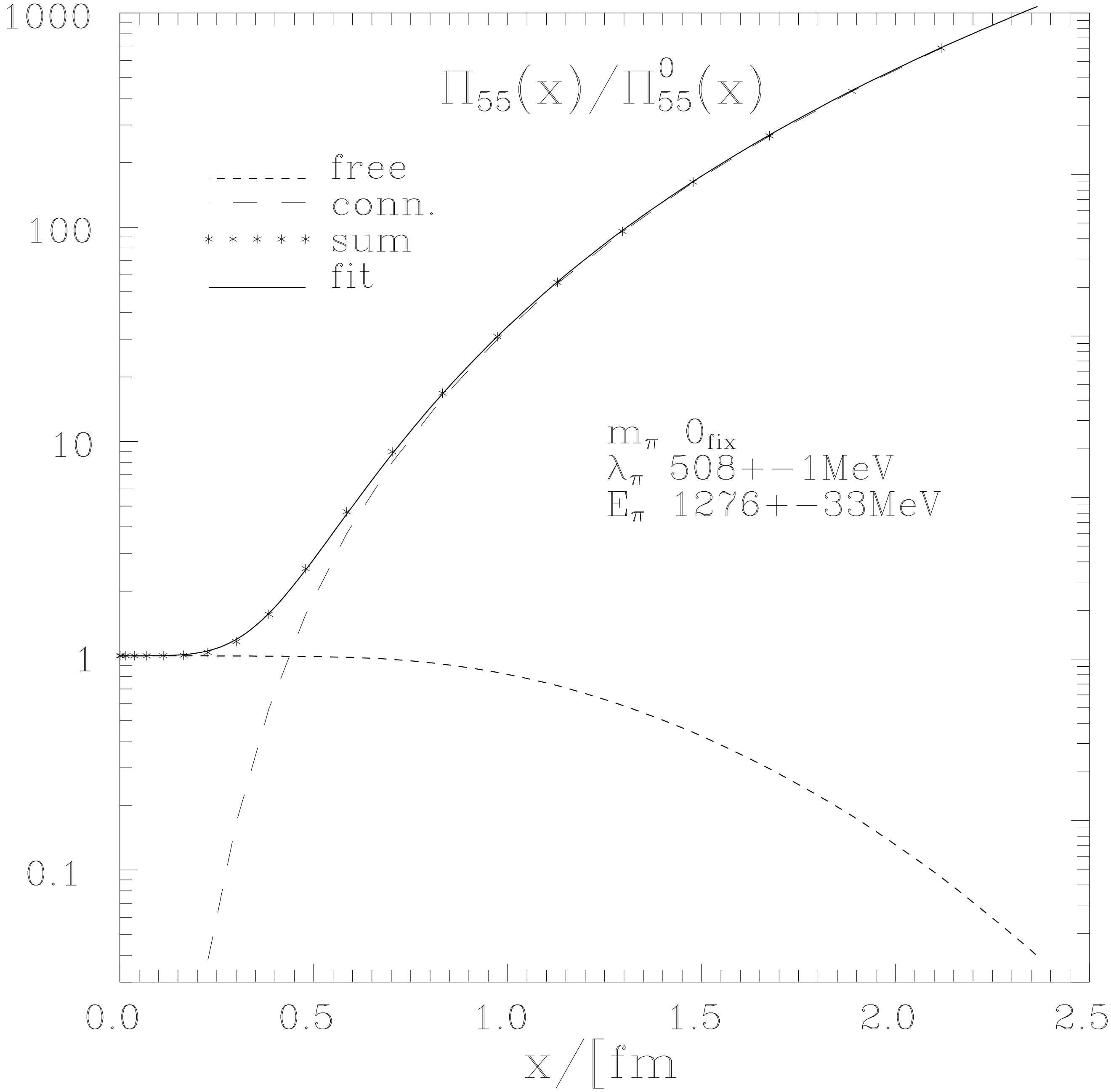

Figure 3: Pseudoscalar triplet correlator normalized to the free massless quark correlator. The pion coupling constant $\lambda_\pi$ and the continuum threshold $E_\pi$ are *fit*ted in order to match the spectral ansatz with the theoretical *sum* of the *free* and the *conn*ected part.

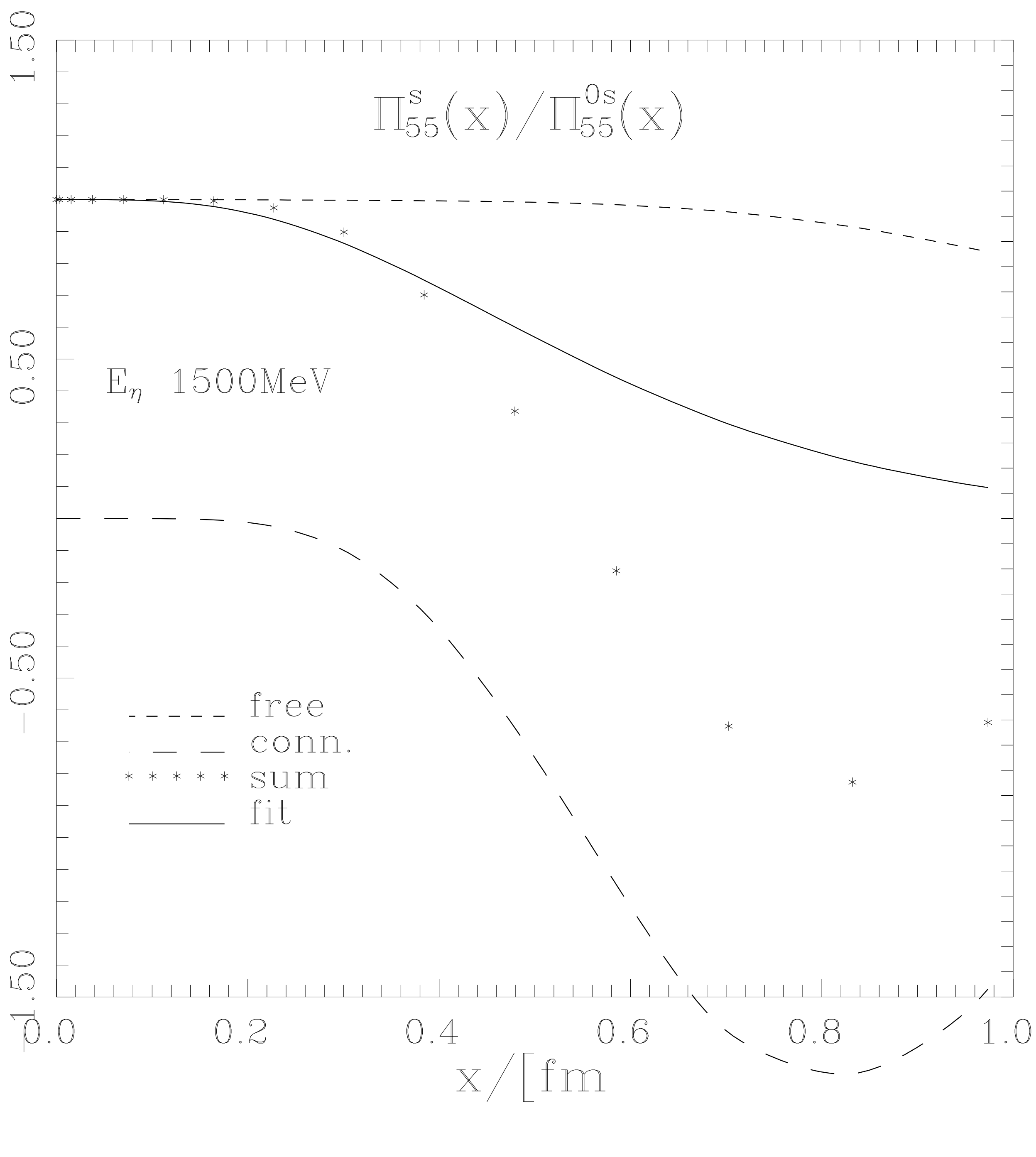

Figure 4: Pseudoscalar singlet correlator normalized to the free massless quark correlator. There is a strong repulsion in this channel and no boundstate is formed. The theoretical curve is compared to a curve obtained from a pure continuum spectrum above $E_{\eta'}$.

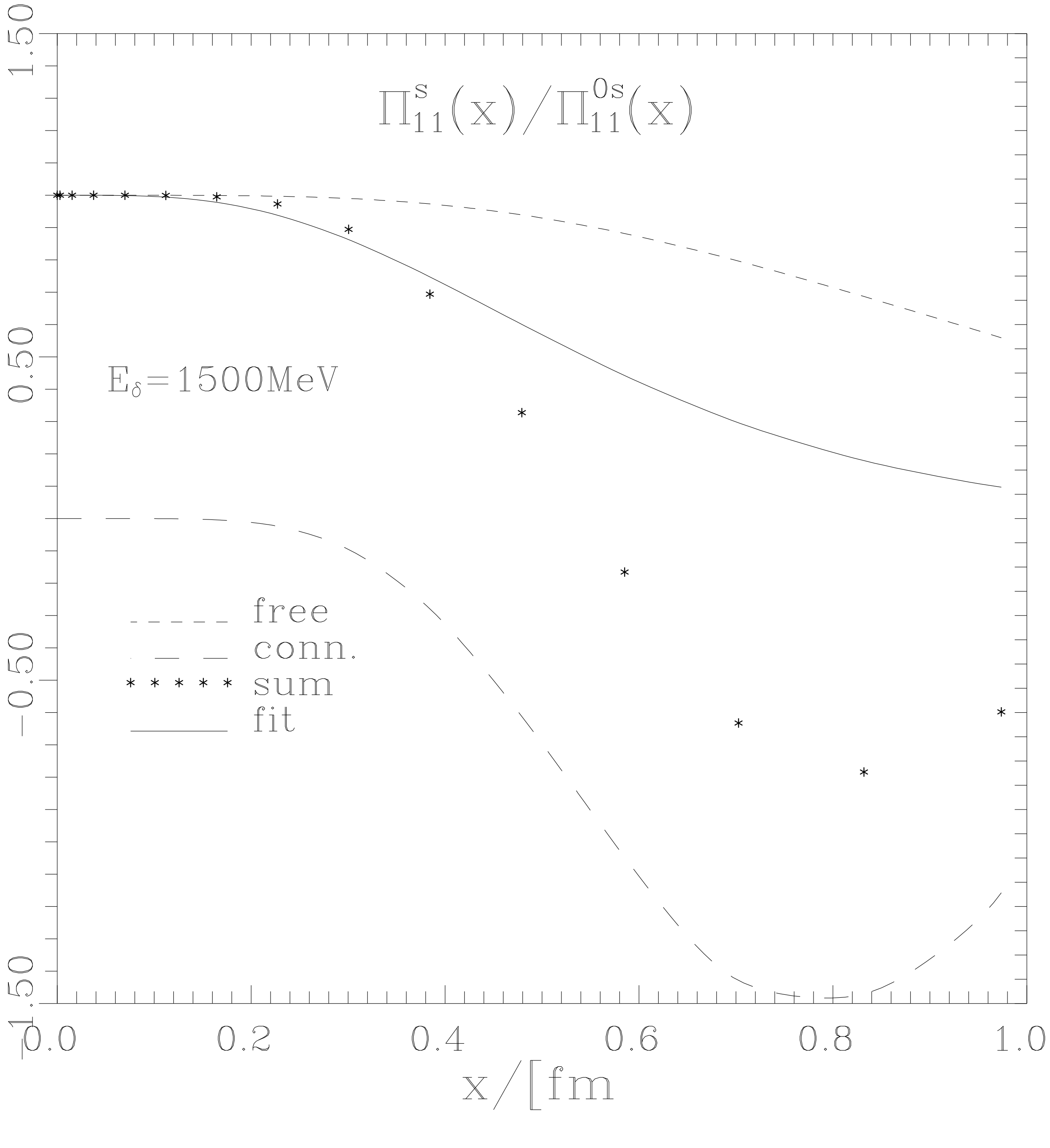

Figure 5: Scalar triplet correlator normalized to the free massless quark correlator. There is a strong repulsion in this channel and no boundstate is formed. The theoretical curve is compared to a curve obtained from a pure continuum spectrum above $E_\delta$.

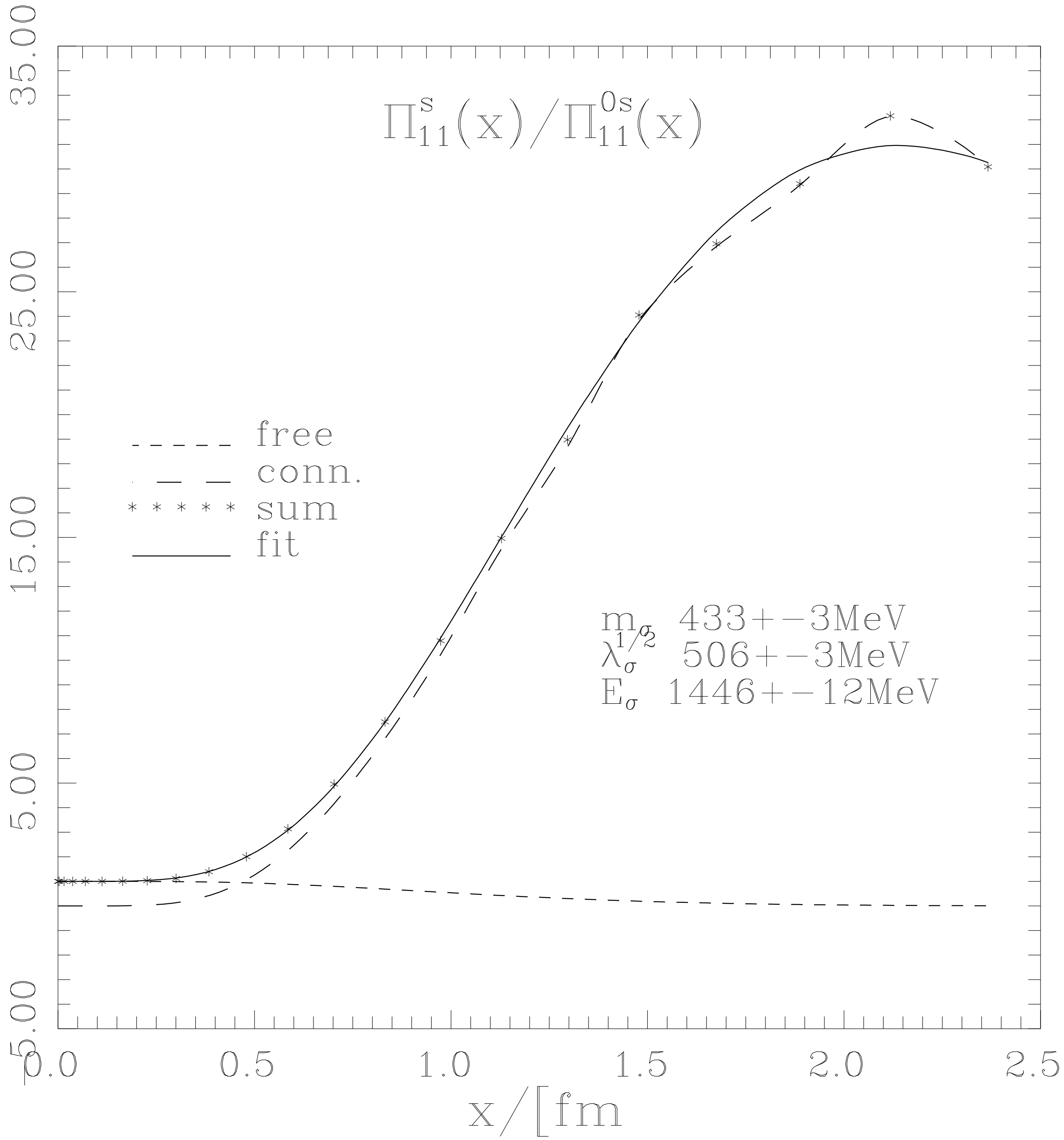

Figure 6: Scalar singlet correlator normalized to the free massless quark correlator. The $\sigma$ mass $m_\sigma$ and coupling $\lambda_\sigma$ and the threshold $E_\sigma$ are obtained from a spectral fit.

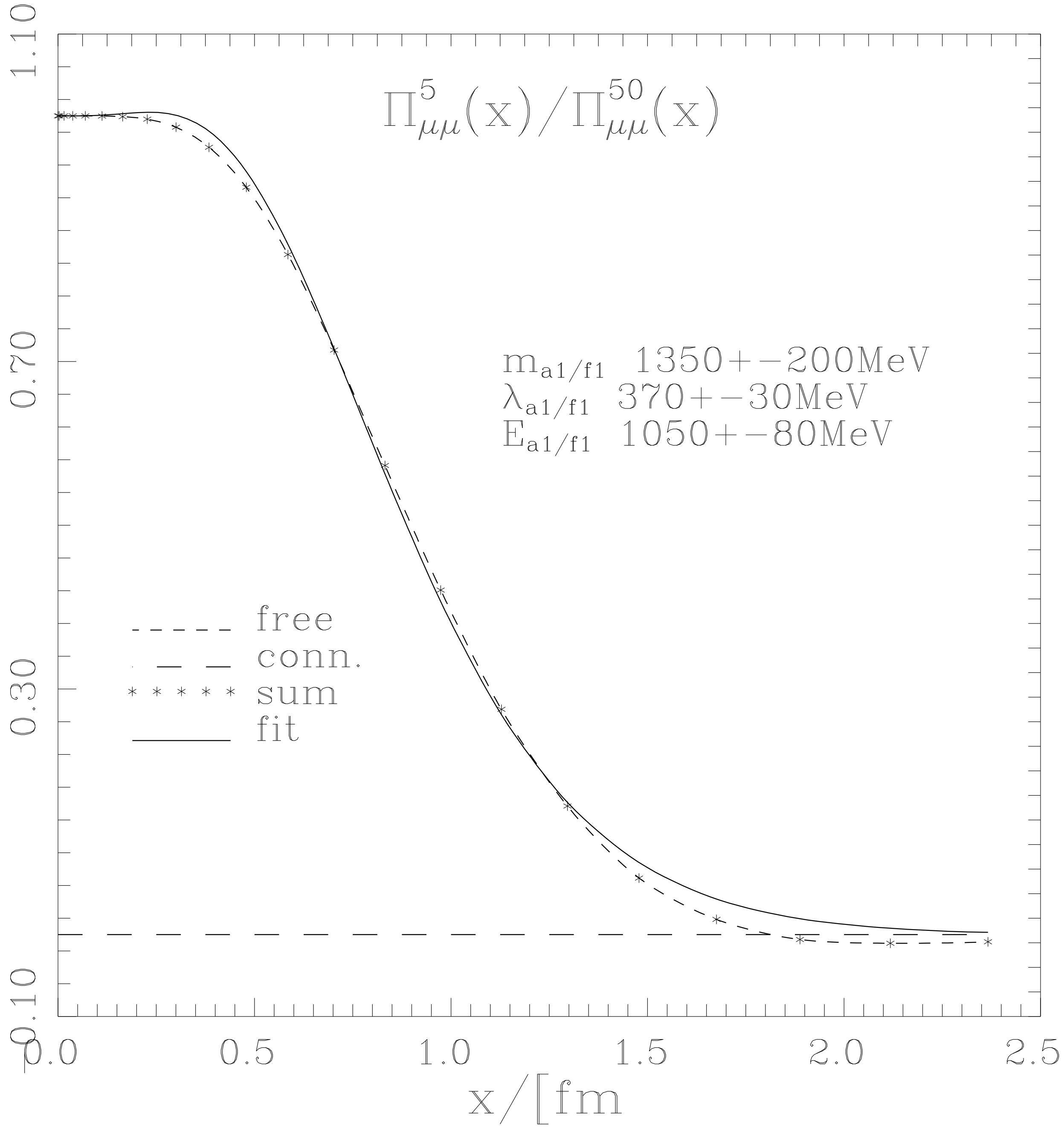

Figure 7: Axial vector correlator normalized to the free massless quark correlator. The triplet and singlet correlator are equal because the connected part has been neglegted. The $a_1$ and $f_1$ mass, coupling and threshold are obtained from a spectral fit.

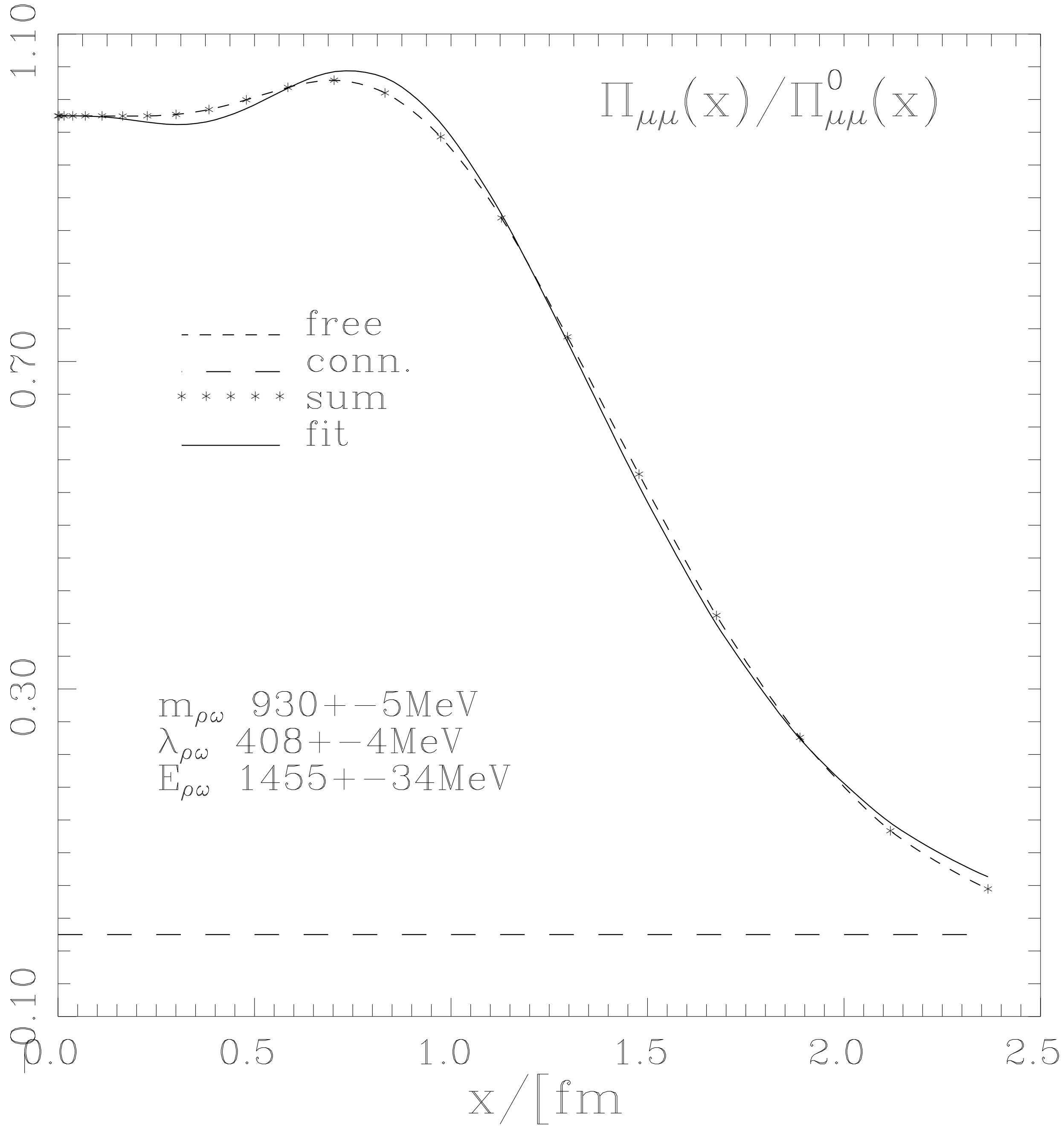

Figure 8: Vector correlator normalized to the free massless quark correlator. The triplet and singlet correlator are equal because the connected part is zero. The $\rho$ and $\omega$ mass, coupling and threshold are obtained from a spectral fit.